\documentclass[11pt]{article}
\usepackage{tabularx}
\usepackage{array}
\usepackage{graphics}
\usepackage{graphicx}
\usepackage{epsfig}                 
\usepackage{amsmath}
\usepackage{amssymb}
\usepackage{psfrag}
\usepackage{color}
\usepackage{longtable}
\usepackage{verbatim}
\usepackage{cite}
\normalsize
\newcommand{\nn}{\nonumber}
\newcommand{\bea}{\begin{eqnarray}}
\newcommand{\eea}{\end{eqnarray}}
\newcommand{\ba}{\begin{array}}
\newcommand{\ea}{\end{array}}
\newcommand{\beq}{\begin{equation}}
\newcommand{\eeq}{\end{equation}}
\newcommand{\no}{\nonumber}

\newcommand{\re}{{Re}}
\newcommand{\im}{{Im}}
\newcommand{\cB}{\mathcal{B}}
\newcommand{\cO}{\mathcal{O}}
\newcommand{\cA}{\mathcal{A}}

\definecolor{Red}{rgb}{1.,0.,0.}
\definecolor{Blue}{rgb}{0.,0.2,1.}

\newcommand{\lsim}{\stackrel{<}{_\sim}}

\def\c{C}

\def\ct{{\c_{10}^{\rm eff}}}
\def\cne{\c_9^{\rm eff}}
\def\cse{\c_7^{\rm eff}}

\def\l{\ell}

\def\d{{\rm d}}

\def\mh{\hat{m}}
\def\mbh{\mh_b}

\def\mvh{\mh_{K^*}}

\def\mlh{\mh_\l}

\def\sh{\hat{s}}

\def\a{{\cal A}}

\def\uh{{\hat{u}}}
\def\la{{\lambda}}

\begin{document}

\thispagestyle{empty}
\begin{flushright}
CERN-PH-TH/2008-161\\
SLAC-PUB-13342\\
\end{flushright}
\vskip 1.5 true cm 

\begin{center}
{\Large\bf Constraints on New Physics in MFV models: \\ [3 mm]
a model-independent analysis of $\Delta F=1$ processes}
  \\ [25 pt]
{\sc  Tobias Hurth${}^{a,b}$, Gino Isidori${}^{c,d}$,  
          Jernej~F.~Kamenik${}^{d,e}$, Federico Mescia${}^{d}$ }
  \\ [25 pt]
{\sl ${}^a$CERN, Dept. of Physics, Theory Division, CH-1211 Geneva, Switzerland \\ [5 pt]  }
{\sl ${}^b$SLAC, Stanford University, Stanford, CA 94309, USA \\ [5 pt]  }
{\sl ${}^c$Scuola Normale Superiore and INFN, Piazza dei Cavalieri 7, 56126 Pisa, Italy \\ [5 pt]  } 
{\sl ${}^d$INFN, Laboratori Nazionali di Frascati, Via E. Fermi 40, 
           I-00044 Frascati, Italy \\ [5 pt] }
{\sl ${}^e$J. Stefan Institute, Jamova 39, P. O. Box 3000, 1001 Ljubljana, Slovenia}
\end{center}

\vskip 2.0 true cm
\begin{abstract}
We analyse the constraints on dimension-six $\Delta F=1$ 
effective operators in models respecting the MFV hypothesis, 
both in the one-Higgs doublet case and in the two-Higgs doublet
scenario with large $\tan\beta$. 
The constraints are derived mainly from the 
$b\to s$ inclusive observables measured at the $B$ factories.
The implications of these bounds in view of improved 
measurements in exclusive and inclusive observables in  
$b\to s \ell^+\ell^-$ and $s\to d \nu\bar\nu$
transitions are discussed. 
\end{abstract}

\newpage

\section{Introduction}

The Standard Model (SM) can be viewed as the renormalizable 
part of an effective field theory, valid up to some still 
undetermined cut-off scale $\Lambda$ above the electroweak 
scale, $ (\sqrt{2}G_F)^{-1/2}\approx 250$~GeV. 
Theoretical arguments based on a natural solution of the 
hierarchy problem suggest that $\Lambda$ should not exceed 
a few TeV. This expectation leads to a paradox when 
combined with the absence of significant deviations 
from the SM in loop-induced flavour-violating 
observables, potentially sensitive to very high 
energy scales. An effective solution to this problem 
is provided by the so-called hypothesis of 
Minimal Flavour Violation~\cite{D'Ambrosio:2002ex}, namely by the 
assumption that the SM Yukawa couplings are the only 
relevant breaking sources of the $SU(3)^5$ flavour 
symmetry~\cite{Chivukula:1987py} of the low-energy effective 
theory.\footnote{~For earlier/alternative definitions of the MFV hypothesis 
see Ref.~\cite{Buras:2000dm,others}.}

This symmetry and symmetry-breaking ansatz
is realised in various explicit extensions 
of the SM, such as supersymmetric models 
{(see e.g.~Ref~\cite{SUSY,SUSYRGE} and~\cite{SUSYR})}
or models with extra dimensions (see e.g.~Ref.~\cite{EDIM}).
However, its main 
virtue is the possibility to perform a general 
analysis of new-physics effects in low-energy 
observables independently of the ultraviolet 
completion of the model.  As shown in Ref.~\cite{D'Ambrosio:2002ex},
the MFV hypothesis allows to build a rather 
predictive effective theory { {in terms 
of SM and  Higgs fields}}. 
The predictions on flavour-violating observables
derived within this effective theory are powerful 
tests of the underlying flavour structure of the 
model: if falsified, these tests would unambiguously signal 
the presence of new symmetry-breaking terms. 

The observables most relevant to test the MFV 
hypothesis and, within this framework, to constrain the 
structure of the effective theory are $\Delta F=2$ and 
$\Delta F=1$ flavour-changing neutral-current (FCNC) processes.
An updated analysis of the $\Delta F=2$ sector, or the 
meson-antimeson mixing amplitudes,   
has been presented recently in Ref.~\cite{UTfit}. 
The goal of this work is a complete analysis 
of the $\Delta F=1$ sector, or the rare-decay 
amplitudes. 

Using the currently available measurements of
$\Delta F=1$ FCNC processes from $b \to s$ and
$s\to d$ transitions (see Table~\ref{tab:obs})
we derive updated bounds on the 
effective scale of new physics within MFV models. 
We consider in particular both the scenario of 
one effective Higgs doublet and the case of 
two Higgs doublets and large $\tan\beta$, where we are 
free to change the  relative normalization of the 
two Yukawa couplings and to decouple the breaking 
of $U(1)_{\rm PQ}$ and $SU(3)^5$ global
symmetries~~\cite{D'Ambrosio:2002ex}.

Having derived the bounds on the effective 
operators from the observables listed in Table~\ref{tab:obs},
we derive a series of predictions for  exclusive 
and inclusive observables in  
$b\to s \ell^+\ell^-$ and $s\to d \nu\bar\nu$
transitions which have not been measured 
so far with high accuracy. On the one hand, these predictions
indicate where to look for 
large new physics effects in the flavour sector, 
even under the pessimistic hypothesis of MFV. On the other 
hand, some of these predictions could 
provide, in the future, a proof of the MFV 
hypothesis: a set of deviations from the SM 
exhibiting the correlation predicted by this
symmetry structure.

The paper is organised as follows: in Section~\ref{sect:MFV}
we review the structure of the effective $\Delta F=1$ 
Hamiltonian under the MFV hypothesis. The theoretical
expressions of the observables analysed, in terms of the 
Wilson coefficients of the effective theory, 
are presented in Section~\ref{sect:obs}. The numerical 
bounds on the scale of new physics and the predictions 
for future measurements are discussed in 
 Section~\ref{sect:bounds} and~\ref{sect:pred}, respectively.
The results are briefly summarised in the Conclusions.

\begin{table}[!t]
\begin{minipage}{\textwidth}
\begin{center}
\begin{tabular}{lll}
  Observable & Experiment &  SM prediction \\  
\hline  
\hline   
$\cB(B \to X_s \gamma)_{[E_{\gamma}>1.6~\mathrm{GeV}]}$ 
   & $(3.52\pm 0.24)\times 10^{-4}$~\cite{HFAG} 
   & $(3.13\pm 0.23)\times 10^{-4}$~\cite{Misiak:2006zs,Misiak:2006ab,Haisch:2008ar} \\
\hline  
$\cB(B \to X_s \ell^+ \ell^-)_{[q^2\in [0.04,1.0]~\mathrm{GeV}^2]}$
   & $\left(0.6 \pm 0.5\right)\times 10^{-6}$ 
   & {$\left(0.8 \pm 0.2\right)\times 10^{-6}$} \\
$\cB(B \to X_s \ell^+ \ell^-)_{[q^2\in [1.0,6.0]~\mathrm{GeV}^2]}$ 
   & $\left(1.6 \pm  0.5\right)\times 10^{-6}$~\cite{Iwasaki:2005sy,Aubert:2004it}\footnote{Here we quote na\"ive averages of the values obtained by the experiments and with symmetrized errors.}
   & {$\left(1.6 \pm 0.1\right)\times 10^{-6}$} 
     ~\cite{Bobeth:1999mk,Asatryan:2001zw,Asatrian:2002va,Ghinculov:2003qd,Bobeth:2003at,Huber:2005ig,Huber:2007vv,Ligeti:2007sn} \\
$\cB(B \to X_s \ell^+ \ell^-)_{[q^2 > 14.4~\mathrm{GeV}^2]}$
   & $\left(4.4 \pm 1.3\right)\times 10^{-7}$  
   & {$\left(2.4 \pm 0.8\right)\times 10^{-7}$}  \\
\hline
$\cB(B_s \to \mu^+ \mu^-)$ 
   & $<5.8 \times 10^{-8}~(95\%~\mathrm{CL})$~\cite{:2007kv} 
   & {$(4.1 \pm 0.8)\times 10^{-9}$}~\cite{Buchalla:1995vs,fbslattice} \\ 
\hline
$\bar A_{FB} (B \to K^* \ell^+ \ell^-)_{[q^2 < 6.25~ \mathrm{GeV}^2]}$ & $0.24^{+0.19}_{-0.24}$ & $-0.01\pm 0.02$ \\
$\bar A_{FB} (B \to K^* \ell^+ \ell^-)_{[q^2 > 10.24~ \mathrm{GeV}^2]}$ &  $0.76^{+ 0.53}_{-0.34}$~\cite{babar:2008ju} & $0.20\pm 0.08$~\cite{Beneke:2001at,Beneke:2004dp,Ali:2006ew,Ball:2004rg} \\
\hline
$\cB(K^+ \to \pi^+ \nu \bar \nu)$ 
   & $\left(14.7^{+13.0}_{-8.9} \right)\times 10^{-11}$~\cite{Adler:2008zz}
   & $(8.6 \pm 0.9)\times 10^{-11}$~\cite{Antonelli:2008jg,Brod:2008ss,Buras:2006gb,Isidori:2005xm,Mescia:2007kn} \\
\end{tabular}
\end{center}
\end{minipage}
\label{tab:obs} \caption{Main observables used to
determine bounds on the MFV dimension-six operators. The SM predictions
are updated according to the most recent determinations of the
SM input values (see Section~\ref{sect:bounds}). }
\end{table}

\section{MFV and $\Delta F = 1 $ processes}
\label{sect:MFV}

Under the MFV hypothesis, the dimension-six 
effective operators relevant to down-type FCNC transitions,
both with one or two Higgs doublets, can be defined 
as follows~\cite{D'Ambrosio:2002ex}:
\beq
\begin{array}{ll}
 {\cal O}_{H1} = i \left( {\bar Q}_L \lambda_{\rm FC} \gamma_\mu
Q_L \right)
 H_U^\dagger D_\mu H_U~, \qquad
&{\cal O}_{H2} = i \left( {\bar Q}_L \lambda_{\rm FC} \tau^a
\gamma_\mu Q_L\right)
 H_U^\dagger \tau^a D_\mu H_U~,\\
 {\cal O}_{G1} =   H_D \left( {\bar Q}_L 
\lambda_{\rm FC} \lambda_d \sigma_{\mu\nu}
  T^a  D_R \right) (g_s G^a_{\mu\nu})~,\qquad
&{\cal O}_{G2} =  \left( {\bar Q}_L \lambda_{\rm FC} \gamma_\mu T^a
Q_L \right) (g_s D_\mu G^a_{\mu\nu})~,\\
  {\cal O}_{F1} = H_D \left( {\bar Q}_L 
\lambda_{\rm FC}  \lambda_d \sigma_{\mu\nu} D_R \right) (e F_{\mu\nu})~,\qquad
&{\cal O}_{F2} = \left( {\bar Q}_L \lambda_{\rm FC} \gamma_\mu Q_L
\right) (e D_\mu
F_{\mu\nu})~,\\
 {\cal O}_{\ell 1} = \left( {\bar Q}_L \lambda_{\rm FC} \gamma_\mu
Q_L \right)
        ({\bar L}_L \gamma_\mu L_L)~, \qquad
&{\cal O}_{\ell 2} = \left( {\bar Q}_L \lambda_{\rm FC} \gamma_\mu
\tau^a Q_L \right)
        ({\bar L}_L \gamma_\mu  \tau^a L_L)~,  \\
 {\cal O}_{\ell 3} = \left( {\bar Q}_L \lambda_{\rm FC} \gamma_\mu
Q_L \right)
        ({\bar E}_R \gamma_\mu E_R)~, &
{\cal O}_{S1} = \left( {\bar Q}_L 
\lambda_{\rm FC} \lambda_d D_R \right) ({\bar E}_R \lambda_{\ell} L_L)~.
\end{array}
\label{eq:basis1}
\eeq
The five fermion fields ($Q_L$,  $D_R$,  $U_R$, $L_L$,  $E_R$) carry a
flavour index ($i=1,\ldots, 3$): since we are interested in FCNC processes of 
down-type quarks, it is convenient to work in the mass-eigenstate 
basis of charged leptons and down-type quarks,
where $\lambda_{d,\ell} = {\rm diag}\{m_{d,\ell}/\langle H_D\rangle\}$. In this basis 
the flavour-changing coupling $(\lambda_{\mathrm{FC}})_{ij}$ can be 
expanded as
\beq
 (\lambda_{\rm FC}) = \left\{
\begin{array}{ll} \left( Y_U Y_U^\dagger \right)_{ij}
\approx \lambda_t^2  V^*_{3i} V_{3j}~ &\qquad i \not= j~, \\
0 &\qquad i = j~, \end{array} \right. 
\label{eq:lfc}
\eeq
in terms of the top-quark Yukawa coupling
($\lambda_t = m_t/\langle H_U\rangle \approx 1$) 
and the CKM matrix ($V_{ij}$).
Note that we have defined the operators linear in the gauge fields 
including  appropriate powers 
of the corresponding gauge couplings (contrary to the original definition of 
Ref.~\cite{D'Ambrosio:2002ex}). Moreover, we have included from the 
beginning the scalar-density operator ${\cal O}_{S1}$, 
which plays a relevant role in the two-Higgs doublet case at large 
$\tan\beta =\langle H_U\rangle/\langle H_D\rangle$.\footnote{~In principle, in 
the two-Higgs doublet case we can consider additional operators 
obtained from those in (\ref{eq:basis1}) by the exchange $H_U  \to H_D$
and/or $\lambda_{\rm FC} \to Y_D Y_D^\dagger \lambda_{\rm FC}$.
However, for all the B physics observables we analyse in this work, the 
effects of these additional operators can be reabsorbed into a 
redefinition of the couplings of the operators in (\ref{eq:basis1}).}
For all the operators linear in $\lambda_d$ we could also 
consider a corresponding set with opposite chirality
($D_R \leftrightarrow Q_L$); 
however, the hierarchical structure of $\lambda_d$ implies
that only one chirality structure is relevant. 
In the following we restrict the attention 
to flavour transitions of the type $j>i$, such as 
the leading chirality structure is the one in
Eq.~(\ref{eq:basis1}).

Before the breaking of the electroweak symmetry, 
we define the MFV effective Lagrangian as
$\mathcal{L}^{\Delta F=1}_{\rm eff-MFV} = (\sum_n a_n {\cal O}_n
)/\Lambda^2$, where $a_i$ are $O(1)$ couplings and $\Lambda$ 
is the effective scale of new physics. After the 
spontaneous breaking of the electroweak symmetry, 
the operators in (\ref{eq:basis1}) are mapped onto 
the standard basis of FCNC operators, defined  by 
\begin{equation}
\mathcal{H}^{\Delta F=1}_{{\rm eff}~[j>i]} ~=~ - \frac{G_{\rm F} \alpha_{\rm em}}{
2\sqrt{2}\pi \sin^2 \theta_{\mathrm{W}}}
 V^*_{3i} V_{3j} \sum_{n} C^{\rm MFV}_n
{\cal Q}_n~+ ~{\rm h.c.}~,
\end{equation}
where
{ {
\beq
\ba{ll}
 {\cal Q}_{7} = \frac{e}{g^{2}}  m_{j} \, \bar d_{i}
 \sigma_{\mu\nu} \left(1+\gamma_5\right)d_{j}
\, F_{\mu\nu}~, 
& \vspace*{0.1cm}
{\cal Q}_{8} = \frac{g_s}{g^{2}}  m_{j}\, 
\bar d_{i} \sigma_{\mu\nu} T^a \left(1+\gamma_5\right) d_{j}
 \, G^a_{\mu\nu}~,  \\
 {\cal Q}_{9} = \bar d_{i} \gamma_\mu \left(1-\gamma_5\right) d_{j}\,   
\sum_\ell ~\bar\ell \gamma_\mu \ell~,  \qquad  
&
\vspace*{0.1cm}
 {\cal Q}_{10}  =  
 \bar d_{i} \gamma_\mu \left(1-\gamma_5\right) d_{j}
\,\sum_\ell ~\bar \ell \gamma_\mu \gamma_5 \ell~,  \\
 {\cal Q}_{\nu\bar\nu} =  \bar d_{i} \gamma_{\mu} \left(1-\gamma_5\right)
 d_{j}
 \,\sum_\nu  ~\bar\nu \gamma_{\mu} \left(1-\gamma_5\right) \nu~,
&
\vspace*{0.1cm}
 {\cal Q}^\ell_{S} =  
\bar d_{i} \left(1+\gamma_5\right) d_{j} \; 
\bar \ell \left(1-\gamma_5\right) \ell~.
\ea
\eeq
}}
Defining 
\beq 
\epsilon_i = \left( \frac{\Lambda_0}{\Lambda}
\right)^2 a_i~, \qquad \Lambda_0 = \frac{\lambda_t \sin^2 \theta_W
m_W}{\alpha_{\rm em} } \approx  2.4 ~{\rm TeV}~, \label{eq:eps_i} 
\eeq
the modified initial conditions of the Wilson coefficients of this 
effective Hamiltonian at the electroweak scale, 
$\delta C_i = C^{\rm MFV}_i(\mu_H)- C^{\rm SM}_i (\mu_H)$, 
are~\cite{D'Ambrosio:2002ex}
\beq
\ba{ll}
  \delta C_{7} = {2g^2}  \epsilon_{F1}~,  
& \delta C_{8} = 2 g^2 \epsilon_{G1}~, \\
  \delta C_{9} = \epsilon_{\ell 1}+\epsilon_{\ell 2}+\epsilon_{\ell 3} 
  - \left[ (1-4\sin^2\theta_W) \epsilon_Z + 2 e^2 \epsilon_{F2} \right]\,,
& \delta C_{10} = \epsilon_Z -\epsilon_{\ell 1} -\epsilon_{\ell 2} +\epsilon_{\ell 3}~,  \\
  \delta C^\ell_{S} = \lambda_{d_j} \lambda_{\ell} ~\epsilon_{S1}~, 
& \delta C_{\nu\bar\nu} = \epsilon_Z + \epsilon_{\ell 1} -\epsilon_{\ell 2}~, 
\ea
\label{eq:C9V}
\eeq
where $\epsilon_Z = (\epsilon_{H1}+\epsilon_{H2})/2$.

At this point it is useful to compare our 
effective $\Delta F=1$ Hamiltonian to those 
adopted in similar analyses in the recent literature. 
Bounds on the scale of new physics and corresponding 
predictions for rare decays in    {MFV scenarios}
have been discussed also by other authors
(see e.g.~Ref.~\cite{Buras:2003td,Bobeth:2005ck,Haisch:2007ia}). 
However, most of the 
recent analyses concentrated on a specific version 
of the general MFV framework~\cite{D'Ambrosio:2002ex}, 
the so-called {\em constrained} MFV (CMFV)~\cite{Buras:2000dm}. 
While  { MFV as} proposed 
in~\cite{D'Ambrosio:2002ex} is an 
hypothesis about the symmetry-breaking structure of the  
$SU(3)^5$ flavour symmetry, 
the CMFV 
contains a further dynamical assumption:
the hypothesis of no new effective dimension-six 
flavour-changing operators 
beside the SM ones (after electroweak symmetry breaking). 
In practice, also this dynamical 
assumption can be related to a symmetry-breaking 
issue: namely the hypothesis about the 
breaking of the $U(1)_{\rm PQ}$ symmetry
of the SM gauge Lagrangian~\cite{D'Ambrosio:2002ex}.
Indeed independently from the value of 
$\tan\beta =\langle H_U\rangle/\langle H_D\rangle$,
in absence of a sizable $U(1)_{\rm PQ}$ breaking
the coefficient of the scalar operator is 
too small to compensate the strong 
$\lambda_{d_j} \lambda_{\ell}$ suppression in Eq.~(\ref{eq:C9V}). 
Since large  $U(1)_{\rm PQ}$ breakings can occur 
in well-motivated extensions of the SM, such as 
supersymmetric models (see e.g.~\cite{Babu:1999hn,Isidori:2001fv,Altmannshofer:2007cs}),  
we make no specific assumptions about its size 
and take into account also the effects of the 
scalar operators.
For the same reason, using the definition of $\lambda_{\rm FC}$ 
in (\ref{eq:lfc}) we remove 
from the operator basis terms contributing to flavour-conserving 
processes induced by the diagonal component of $Y_U Y_U^{\dagger}$. 
At large $\tan\beta$ flavour-diagonal terms could 
receive additional contributions of the type  $Y_D Y_D^{\dagger}$, 
therefore we cannot use them to bound the flavour-changing terms. 
This does not occur in the CMFV, where $Z\to b\bar b$ provide
a useful constraint on  flavour-changing operators~\cite{Haisch:2007ia}.

\subsection{Effective weak Hamiltonian at  $\mu=m_b$}
When computing $\Delta F=1$ observables, 
we need to take into account also the contributions 
of four-quark operators. Following the notation of 
Ref.~\cite{Bobeth:1999mk}, we write the complete 
effective Hamiltonian relevant to $b\to s$ transitions as
\bea 
{\cal H}^{ b\to
s}_{\rm eff} &=& -\frac{4 G_F}{\sqrt{2}}
 [  V^*_{us} V_{ub} (C^c_1 P^u_1 + C^c_2 P^u_2)
  + V^*_{cs} V_{cb} (C^c_1 P^c_1 + C^c_2 P^c_2)]
\nonumber\\
&-& \frac{4 G_F}{\sqrt{2}} \left\{\sum_{i=3}^{10} [(V^*_{us} V_{ub}
+ V^*_{cs} V_{cb}) C^c_i \; + \; V^*_{ts} V_{tb} C^t_i] P_i +
V^*_{ts} V_{tb}
C^\ell_{0} P^\ell_{0}\right\}~+~{\rm h.c.}\,, \label{Leff} 
\label{eq:newHeff}
\eea 
where
\beq
\begin{array}{ll}
P^u_1 =  (\bar{s}_L \gamma_{\mu} T^a u_L) (\bar{u}_L \gamma^{\mu}
T^a b_L)~, &
\vspace*{0.3cm}
P_5 =  (\bar{s}_L \gamma_{\mu_1}
                   \gamma_{\mu_2}
                   \gamma_{\mu_3}    b_L)\sum_q (\bar{q} \gamma^{\mu_1}
                                                         \gamma^{\mu_2}
                                                         \gamma^{\mu_3}     q)~,
 \\
P^u_2 =  (\bar{s}_L \gamma_{\mu}     u_L) (\bar{u}_L \gamma^{\mu}
b_L)~, & \vspace*{0.3cm}
P_6 =  (\bar{s}_L \gamma_{\mu_1}
                   \gamma_{\mu_2}
                   \gamma_{\mu_3} T^a b_L)\sum_q (\bar{q} \gamma^{\mu_1}
                                                          \gamma^{\mu_2}
                                                          \gamma^{\mu_3} T^a q)~,\\
P^c_1 =  (\bar{s}_L \gamma_{\mu} T^a c_L) (\bar{c}_L \gamma^{\mu}
T^a b_L)~,
& \vspace*{0.3cm}
P_7  =   \frac{e}{g_s^2} m_b (\bar{s}_L \sigma^{\mu \nu}     b_R)
F_{\mu \nu}~,\\
P^c_2 =  (\bar{s}_L \gamma_{\mu}     c_L) (\bar{c}_L \gamma^{\mu}
b_L)~,
&\vspace*{0.3cm}
P_8  =   \frac{1}{g_s} m_b (\bar{s}_L \sigma^{\mu \nu} T^a b_R)
G_{\mu \nu}^a~,\\
P_3 =  (\bar{s}_L \gamma_{\mu}     b_L) \sum_q (\bar{q}\gamma^{\mu}
q)~,
&\vspace*{0.3cm}
P_{10} =  \frac{e^2}{g_s^2} (\bar{s}_L \gamma_{\mu} b_L) \sum_\ell
                             (\bar{\ell} \gamma^{\mu} \gamma_5 \ell)~,\\
P_4 =  (\bar{s}_L \gamma_{\mu} T^a b_L) \sum_q (\bar{q}\gamma^{\mu}
T^a q)~,
&\vspace*{0.3cm}
P_9  =   \frac{e^2}{g_s^2} (\bar{s}_L \gamma_{\mu} b_L) \sum_\ell
 (\bar{\ell}\gamma^{\mu} \ell)~. \\
\end{array}
\eeq 
With respect to the SM literature we also add the 
scalar-density operator with right-handed $b$ quark,
\beq 
P^\ell_{0} = \frac{e^2}{16\pi^2} (\bar s_L b_R) (\bar \ell_R \ell_L)~,
\eeq 
which plays an important role in the large $\tan\beta$ regime.
This operator is present also in the SM, but it is usually neglected 
because of the of strong suppression of its Wilson coefficients.

In principle we should consider new-physics effects both 
in the four-quark ($P_{1-6}$) and in the FCNC operators. 
However, four-quark operators receive a large SM contribution 
which is naturally much larger with respect to the new-physics 
one in the MFV scenario. Moreover four-quark operators do not 
contribute at the tree-level to the observables we are 
considering (FCNC processes). As a result, it is a good 
approximation to consider new-physics effects only 
in the leading FCNC operators:\footnote{~Even including explicit 
new-physics contributions to $C_{1-6}$, to an excellent 
accuracy these could reabsorbed into the Wilson 
coefficients of the FCNC operators.}
\begin{eqnarray}
C^t_7 (\mu_H) &=& C^t_{7,SM} (\mu_H) + \frac{\alpha_s}
{4\pi} \delta C_{7}~, \no \\
C^t_8 (\mu_H) &=& C^t_{8,SM} (\mu_H) + \frac{\alpha_s}
{4\pi} \delta C_{8}~, \no \\
C^t_9 (\mu_H) &=& C^t_{9,SM} (\mu_H) + \frac{\alpha_s}
{4\pi\sin^2 \theta_{\mathrm W}}  \delta C_{9} ~, \\
C^t_{10} (\mu_H) &=& C^t_{10,SM} (\mu_H) + \frac{\alpha_s}
{4\pi\sin^2 \theta_{\mathrm W}}  \delta C_{10} ~, \no \\
C^\ell_{0} (\mu_H) &=& \frac{2} { \sin^2 \theta_{\mathrm W}}
\delta C^{\ell}_{S}  ~, \no
\end{eqnarray}
with the $\delta C_i$ given in Eq.~(\ref{eq:C9V}).
The initial conditions for the SM terms are obtained 
by a NNLO matching of the effective Wilson coefficients 
at the high scale $\mu_H \in [80,160]~\mathrm{GeV}$, 
using  CKM unitarity to get rid 
of $V^*_{cb} V_{cs}/V^*_{tb} V_{ts}$~\cite{Bobeth:1999mk}.

To a good approximation, in all the considered low-energy observables 
$\delta C_7$ and $\delta C_8$ appear  in a fixed linear combination:
\beq
\delta C_7 + 0.3~\delta C_8~.
\label{eq:C7C8rel}
\eeq
For this reason, in the following we set $\delta C_8=0$ and treat
only $\delta C_7$ as independent fit parameter (avoiding the  
flat direction in the parameter space determined by Eq.~(\ref{eq:C7C8rel})).
The bounds on  $\delta C_7$ thus obtained should therefore be 
interpreted as bounds on the $\delta C_7$--$\delta C_8$
combination in Eq.~(\ref{eq:C7C8rel}).

The Wilson coefficients are evolved from the high scale down to 
$\mu_b =O(m_b)$ using SM renormalization 
group equations at the  NNLO accuracy~\cite{Gambino:2003zm,Gorbahn:2004my}. 
At the low scale it is convenient to define the effective
coefficients~\cite{Asatrian:2002va} 
\bea 
A_7 &=& \frac{4
\pi}{\alpha_s(\mu_b)} C_7(\mu_b) - \frac{1}{3}  C_3(\mu_b) -
\frac{4}{9} C_4(\mu_b) - \frac{20}{3}  C_5(\mu_b) - \frac{80}{9}
C_6(\mu_b), \\
 A_9 &=& \frac{4 \pi}{\alpha_s(\mu_b)} C_9(\mu_b)
+ \sum_{i=1}^{6} C_i(\mu_b) \gamma_{i9}^{(0)}  \ln \frac{m_b}{\mu}
 + \frac{4}{3} C_3(\mu_b)
+ \frac{64}{9}  C_5(\mu_b) + \frac{64}{27}  C_6(\mu_b), \\
 A_{10} &=&
\frac{4 \pi}{\alpha_s(\mu_b)} C_{10}(\mu_b), \\
T_9 &=& \frac{4}{3} C_1(\mu_b) + C_2(\mu_b)+ 6 C_3(\mu_b)+ 60
C_5(\mu_b) ,\\
 U_9 &=& - \frac{7}{2} C_3(\mu_b) - \frac{2}{3}
C_4(\mu_b) -38  C_5(\mu_b) - \frac{32}{3} C_6(\mu_b), \\
W_9 &=& - \frac{1}{2}  C_3(\mu_b) - \frac{2}{3} C_4(\mu_b) -8
C_5(\mu_b) - \frac{32}{3}  C_6(\mu_b)\,, 
\eea 
where $C_i(\mu) = C^t_i(\mu) - C^c_i(\mu)$, 
$\gamma_{i9}^{(0)}$ is given in~\cite{Bobeth:1999mk}
and we have neglected the tiny $u$-quark loop contribution to $A_9$.
The reference input values used in this procedure are collected in
Table~\ref{table:setup}.

\begin{table}
\begin{center}
\begin{tabular}{ll}
  Quantity & Value \\
  \hline\hline
  $\sin^2 \theta_W$ & $0.2312$ \\
  $\alpha_s(m_Z)$ & $0.1187$ \\
  $\alpha_{\rm em}(m_Z)$ & $1/127.8$ \\
  \hline
  $m_b^{\overline{MS}}(m_b)$ & $4.2~\mathrm{GeV}$ \\
  $m_t^{\overline{MS}}(m_t)$ & $165~\mathrm{GeV}$ \\
\end{tabular}
\caption{Reference input values used in the
matching and in the RGE evolution of the Wilson coefficients.
\label{table:setup}}
\end{center}
\end{table}

Beside $B$ physics, we are interested also in 
$s \to d$ transitions and particularly in 
the rare decays $K\to \pi \nu\bar\nu$. 
In the MFV framework these are described by the 
following simple effective Hamiltonian
\beq
\mathcal{H}^{s\to d}_{\rm eff}=\frac{G_{F}\alpha_{\rm em}\left( m_{Z}\right)
}{\sqrt{2}}{{\sum_{\ell=e,\mu,\tau}}}\left(  \frac{y_{\nu}}{2\pi\sin^{2}\theta_{W}}P_{\nu\bar{\nu}%
}\right)  +\mathrm{h.c.} \;,
\eeq
where
\bea
P_{\nu\bar{\nu}}&=&\left(  \bar{s}\gamma_{\mu}d\right)  \left(
\bar{\nu }_{\ell}\gamma^{\mu}\left(  1-\gamma_{5}\right)
\nu_{\ell}\right)~, \\
y_{\nu} &=&{\frac{1}{\vert V_{us}\vert}}\left(
\lambda_{t} (X_{t} + \delta C_{\nu\bar\nu})+{Re}\lambda_{c}\widetilde{P}_{u,c}\right)\;,
\label{NUCoeff}%
\eea
with $\lambda_{q}=V_{qs}^{\ast}V_{qd}$, 
$X_{t}=1.464\pm0.041$, $\widetilde{P}_{u,c}=(0.2248)^4\,P_{u,c}$~\cite{Antonelli:2008jg} and 
$P_{u,c}=0.41\pm0.04$~\cite{Brod:2008ss,Buras:2006gb,Isidori:2005xm}.

\section{Observables}
\label{sect:obs}

The observables we are interested in, 
either to derive bounds from the existing measurements 
or to obtain predictions for future measurements, 
are the differential 
decay distributions of $B \to X_s \ell^+\ell^-$
and $B \to (K^*,K) \ell^+ \ell^-$ decays, 
and the integrated rates of $B_{s,d} \to \ell^+ \ell^-$,
$B\to X_s \gamma$, $B \to K^{(*)} \nu \bar \nu$ and  $K \to \pi\nu\bar\nu$.
In the following we analyse the theoretical
expressions of these observables in terms of the 
(non-standard) Wilson coefficients 
of the MFV effective theory.

\subsection{$B \to X_s \ell^+\ell^-$}

In order to minimise the theoretical uncertainty, 
we normalise all the observables to the corresponding 
SM predictions. 
The numerical values for the SM predictions are computed 
at full NNLO accuracy, applying also QED corrections, 
following the analysis of Ref.~\cite{Huber:2005ig,Huber:2007vv}, 
and  non-perturbative $1/m_b$ corrections~\cite{Ghinculov:2003qd,Ligeti:2007sn}.
On the other hand, since full NNLO expressions for the 
non-standard operator basis in (\ref{eq:newHeff}) are not available, 
the relative deviations from the SM are
computed at NLO accuracy (with partial inclusion of NNLO 
corrections, as discussed below).
Given the overall good agreement between data and SM predictions,
and the corresponding smallness of non-standard effects,
this procedure minimise the theoretical uncertainty.

Following the standard notations of $B \to X_s \ell^+\ell^-$
analyses within the SM, we define the effective coefficients 
\begin{subequations}
\begin{eqnarray}
\widetilde{C}_7^{\rm eff} &=& \left[1+ \frac{\alpha_s}{\pi}
\omega_7(\hat s) \right] A_7+\ldots,\\
\widetilde{C}_9^{\rm eff} &=& \left[1+ \frac{\alpha_s}{\pi}
\omega_9(\hat s) \right] [A_9+ T_9  h (\hat m_c^2, \hat{s})+U_9 h
(1,\hat{s})+
    W_9  h (0,\hat{s})]+\ldots,\\
\widetilde{C}_{10}^{\rm eff} &=& \left[1+ \frac{\alpha_s}{\pi}
\omega_9(\hat s) \right] A_{10}+\ldots,
\end{eqnarray}
\end{subequations}
where all quantities with a hat are normalized to the $b$ 
quark pole mass ($\hat x=x/m_b$). 
The leading-order gluon 
bremmsstrahlung corrections $\omega_i(\hat s)$ and the 
loop function $ h (x, y)$ can be found in~\cite{Ghinculov:2003qd,Asatrian:2002va}, while the dots denote additional perturbative $\alpha_s$ and electro-magnetic corrections, as well as non-perturbative power corrections, which we only consider for the SM normalization.
The Wilson coefficients are evolved down 
to the low scale $\mu_b=2.5~\mathrm{GeV}$, using the input values 
in Table~\ref{table:setup}. 
As shown in \cite{Ghinculov:2003qd},  
using such a low renormalisation scale minimise the 
NNLO QCD corrections 
to the $B \to X_s \ell^+\ell^-$ differential rate.
An important difference with respect to the SM case is the presence 
of the $P_{0}^{\ell}$ operator. To the level of accuracy we are working at,
its inclusion is straightforward, since it is renormalised only in a multiplicative way.

We are now ready to compute the inclusive decay spectra.
Neglecting the strange quark mass while keeping the full 
dependence on the lepton mass, we get 
for the unnormalized forward-backward assymmetry and the rate
\bea
\frac{\mathrm{d}A_{FB}}{\mathrm{d}\hat s} &=& 
\frac{G_F^2   \alpha_{\mathrm{em}}^2
   m_b^5}{256 \pi ^5}\left| V_{tb}^{} V_{ts}^\ast  \right|^2 (1-\hat{s})^2  \left(1-\frac{4 \hat{m}_\ell^2}{\hat{s}^2}\right) \nn\\
&&\times    \left\{
-2
   \re\left(\widetilde C_9^{\rm eff}
   \widetilde C_{10}^{{\rm eff} *}
   \right)
   \hat{s}-4
   \re\left(\widetilde C_7^{\rm eff}
   \widetilde C_{10}^{{\rm eff} *}
   \right)
+ 
   \re\left(\widetilde C_9^{\rm eff}
   C_{0}^{\ell *}
   \right) \hat{m}_\ell+{2
   \re\left(\widetilde C_7^{\rm eff}
   C_{0}^{\ell *}
   \right)
   \hat{m}_\ell}\right\}\,,\nn\\
&&\\
 \frac{d \Gamma }{d \hat{s}} &=& 
\frac{G_F^2   \alpha_{\mathrm{em}}^2 m_{b}^5 }{768 \pi^5} 
      \left| V_{tb}^{} V_{ts}^\ast  \right|^2   
(1-\hat{s})^2 \sqrt{1-\frac{4 \hat{m}_\ell^2}{\hat{s}^2}}\nn\\
&\times & \Bigg\{
\Bigg[12 Re ( \widetilde{C}_7^{\rm eff} \widetilde{C}_9^{\rm eff\ast}) +
\frac{4 |\widetilde{C}_7^{\rm eff}|^2 (2+\hat{s})}{\hat{s}}\Bigg]
\Bigg(1+\frac{2 \hat{m}_\ell^2}{\hat{s}}\Bigg)
+ 6 \hat{m}_\ell^2(|\widetilde{C}_9^{\rm eff}|^2 - |\widetilde{C}_{10}^{\rm eff}|^2)\nn\\
&+& (|\widetilde{C}_9^{\rm eff}|^2 + |\widetilde{C}_{10}^{\rm eff}|^2)\Bigg[1+2 \hat{s} +
  \frac{2\hat{m}_\ell^2 (1-\hat{s}) }{\hat{s}}\Bigg] 
+ \frac{3}{4} \hat{s}\Bigg[\Bigg(1-\frac{4
  \hat{m}_\ell^2}{\hat{s}}\Bigg) |C_0^{\ell}|^2 \Bigg]\nn\\
&+& 3 \hat{m}_\ell Re( C_0^{\ell} \widetilde{C}_{10}^{\rm eff \ast})\Bigg\}\,.
\end{eqnarray}
As anticipated, these expressions are used only to evaluate 
the relative impact of new physics. 
Altogether, the estimated theoretical errors
for the integrated inclusive rates reported 
in Table~\ref{tab:obs} are around 
$10\% - 30\%$.
In the numerical analysis 
this error is added in quadrature to the experimental one, 
which, at present, provides the largely dominant source of uncertainty.\footnote{~There 
are two sources of theoretical uncertainties which, at present, 
are difficult to estimate: i) non-perturbative power corrections 
of order $\alpha_s \Lambda / m_b$; ii) radiative corrections 
associated to the soft-photon
experimental cuts applied in the experiments. 
In the first case the effect is estimated to be 
of $O(5 \%)$ by simple dimensional counting~\cite{BIR}
(more sophisticated estimates using the vacuum insertion 
method in the context of $B \rightarrow X_s \gamma$
confirm this naive estimate~\cite{Neubertlee}).
As far as radiative corrections are concerned, 
the theoretical predictions include a correction 
due to collinear logs \cite{Huber:2007vv} 
which does not correspond to the treatment of soft and 
collinear photons performed so far  by the experiments 
\cite{Hubernew}. We guesstimate this 
mismatch as a $10\%$ uncertainty in the high-$q^2$ region 
and $5\%$ error in the low-$q^2$ region.
Given the large experimental uncertainties, which are still
dominant, the impact of these additional theory errors
turns out to be negligible. However, 
we stress that in the future a reduction 
of the theory uncertainties can be obtained only with a 
proper treatment of soft-photon corrections. }

\subsection{$B \to (K^*,K) \ell^+ \ell^-$, $B \to (K^*,K) \nu \bar\nu$}
\label{sect:exclusive}

Exclusive modes are affected by a substantially larger theoretical 
uncertainty with respect to the inclusive ones. However, they are
experimentally easier and allow to probe or constrain combinations 
of Wilson coefficients which are not accessibe (or hardly accessible) 
in the inclusive modes. The most interesting observables 
for our analysis are the forward-backward asymmetry (FBA) 
in $B \to K^* \ell^+ \ell^-$ and the 
lepton universality ratio in $B \to K^{(*)} \ell^+ \ell^-$. In addtion, interesting predictions can be derived for the $B \to (K^*,K) \nu \bar\nu$ decay rates.

In the $B \to K^* \ell^+ \ell^-$ case we need to 
introduce seven independent hadronic 
form factors ($V$, $A_{0-2}$, $T_{1-3}$~\cite{Ball:2001fp,Ball:2004rg}). 
Following Refs.~\cite{Ali:1999mm,Hiller:2003js} we define the following set of   
auxiliary variables:
\bea
  A(\sh) & = & \frac{2}{1 + \mvh} \cne(\sh) V(\sh) 
         + \frac{4 \mbh}{\sh} \cse T_1(\sh) \; , \no\\ 
  B(\sh) & = & (1 + \mvh) \left[ \cne(\sh) A_1(\sh) 
         + \frac{2 \mbh}{\sh} (1 - \mvh) \cse T_2(\sh) \right] \; , \no\\
  C(\sh) & = & \frac{1}{1 - \mvh^2} \left[ 
         (1 - \mvh) \cne(\sh) A_2(\sh) 
         + 2 \mbh \cse \left( 
           T_3(\sh) + \frac{1 - \mvh^2}{\sh} T_2(\sh) \right) \right] \; , 
\no \\
  D(\sh) & = & \frac{1}{\sh} \left[ \cne(\sh) \left(
       (1 + \mvh) A_1(\sh) - (1 - \mvh) A_2(\sh) - 2 \mvh A_0(\sh) \right) 
      - 2 \mbh \cse T_3(\sh) \right] \; , \no\\
  E(\sh) & = & \frac{2}{1 + \mvh} \ct V(\sh) \; , \no\\
  F(\sh) & = & (1 + \mvh) \ct A_1(\sh) \; , \no\\
  G(\sh) & = & \frac{1}{1 + \mvh} \ct A_2(\sh) \; , \no\\
  H(\sh) & = & \frac{1}{\sh} \ct \left[
       (1 + \mvh) A_1(\sh) - (1 - \mvh) A_2(\sh) - 2 \mvh A_0(\sh) \right] \; , \no \\
  X(\sh) & = & -\frac{ m_{K^*}}{m_b} A_0(\sh) C_{0}^\ell \, ,
\label{eq:aux2}
\end{eqnarray}
where now all quantities with a hat are normalized to the physical 
meson mass (and bremmstrahulng corrections are omitted from the
effective Wilson coefficients without the tilde). Also in this case the above 
expressions are used only to evaluate deviations from the SM.
As far as the SM predictions are concerned, in this case 
the most accurate results are obtained by means of 
QCD factorisation and SCET~\cite{Beneke:2001at,Beneke:2004dp,Ali:2006ew,Hill:2004if,Egede:2008uy}.
However, here we adopt a conservative point of view and
evaluate the SM predictions and the corresponding 
errors by means of the parametrisation and QCD SR calculations of Ref.~\cite{Ball:2004rg}. Then we enlarge the theoretical errors by the differences between our SM predictions and known QCD factorization and SCET results~\cite{Beneke:2001at,Beneke:2004dp,Ali:2006ew}.
In practice, given the large experimental errors,
this choice has almost no influence on the 
constraints derived. 

Using the auxiliary variables in Eq.~(\ref{eq:aux2}),
the un-normalised FBA then reads
\begin{eqnarray}
 \frac{\d \a_{\rm FB}}{\d \sh} &=&
\frac{G_F^2 \, \alpha^2_{\mathrm{em}} \, m_B^5}{2^{10} \pi^5}   \left| V_{ts}^\ast V_{tb} \right|^2 \,  \uh(\sh)^2
\times \no \\
&&  \times 
      \left\{ \sh {Re}(BE^\ast) + \sh {Re}(AF^\ast)- \frac{\mlh}{\mvh^2} \left[ \la {Re}(X C^\ast) - (1-\mvh^2 -\sh ) {Re}(X B^\ast) \right] \right\}, \qquad
  \label{eq:dfbabvllex}
\end{eqnarray}
where
\begin{eqnarray}
  \uh(\sh) & =& \sqrt{\la (1-4 \frac{\mlh^2}{\sh})}  \; , \\
 \la & \equiv& \la(1,\hat{m}_{K^*}^2,\sh) =
1+\hat{m}_{K^*}^4+\sh^2-2 \sh-2 \hat{m}_{K^*}^2(1+\sh) \; .
\end{eqnarray}
Similarly the decay rate can be written as 
\begin{eqnarray}
&&  \frac{d \Gamma }{d \hat s} =  
  \frac{G_F^2 \, \alpha_{\rm em}^2 \, m_B^5}{2^{10} \pi^5} 
      \left| V_{ts}^\ast \, V_{tb} \right|^2 
  \times\Bigg\{ 
  \frac{|A|^2}{3} \la 
   \left(\sh  + \mlh^2 \right) 
  + \frac{|E|^2}{3} \la \left(\sh - \mlh^2  \right)   
        \Bigg.
        \nonumber \\
  &&\qquad  \Bigg.
  + \frac{1}{4 \mvh^2} \left[ 
  |B|^2 \left( \la - \uh(\sh)^2/3 + 8 \mvh^2 (\sh +2 \mlh^2 ) \right)
  + |F|^2 \left( \la - \uh(\sh)^2/3 + 8 \mvh^2 (\sh -4 \mlh^2) \right) \right] 
        \Bigg.
        \nonumber \\
  &&\qquad + \frac{\la}{\mvh^2} \left[ \sh |X|^2 + 2 \mlh \la ( { Re}(X F^\ast) - \sh { Re}(X H^\ast) - (1-\mvh^2 ) { Re}(X G^\ast)  ) \right] 
		\nonumber \\
  &&\qquad \Bigg.
  + \frac{\la}{4 \mvh^2} \left[ |C|^2 (\la -\uh(\sh)^2/3) 
    + |G|^2 \left( \la -\uh(\sh)^2/3 + 4 \mlh^2 (2 + 2 \mvh^2 - \sh) \right) \right]
        \Bigg.
        \nonumber \\
  &&\qquad \Bigg.
  - \frac{1}{2 \mvh^2} \left[ 
  { Re}(BC^\ast +FG^\ast) (1 - \mvh^2 - \sh)(\la - \uh(\sh)^2/3) 
  +  4 \mlh^2 {Re}(FG^\ast) \la  \right]
        \Bigg.
        \nonumber \\
  &&\qquad \Bigg.
  - 2 \frac{\mlh^2}{\mvh^2} \la \left[ 
    {Re}(FH^\ast)
    - {Re}(GH^\ast) (1 - \mvh^2) \right] 
 + |H|^2 \frac{\mlh^2}{\mvh^2} \sh \la
  \Bigg\} \; .
   \label{eq:ddwbvll}
\end{eqnarray}

\medskip
\noindent 
In the $B\to K\ell^+\ell^-$ case we only need three
form factors ($f_+$, $f_0$ and $f_T$)~\cite{Ball:2004ye}. Again we define a set of auxiliary variables~\cite{Ali:1999mm,Hiller:2003js}
\bea
\label{eq:aux1}
  A^{\prime} &=&  {C}_9^{\mathrm{eff}}(\hat{s}) f_+(\hat{s}) 
         + \frac{2 \hat{m}_b}{1 + \hat{m}_K} {C}_7^{\mathrm{eff}} f_T(\hat{s}),\nonumber\\
C^{\prime} &=&  {C}^{\mathrm{eff}}_{10}  f_+(\hat{s}),\nonumber\\
D^{\prime} & =&   \frac{1-\hat{m}_K^2}{\hat{s}}{C}^{\mathrm{eff}}_{10}[f_0(\hat{s}) - f_+(\hat{s})],\nonumber\\
X^{\prime}  &=& \frac{1-\hat{m}_K^2}{2\hat{m}_b} C_{0}^{\ell} f_0(\hat{s}) ,
\eea
in terms of which
\begin{eqnarray}
\lefteqn{\frac{d \Gamma }{d \hat{s}}  =  
  \frac{G_F^2  \alpha_{\mathrm{em}}^2  m_B^5}{2^{10} \pi^5} 
      \left| V_{tb}^{} V_{ts}^\ast  \right|^2  \hat{u}(\hat{s}) 
\Bigg\{ 
(|A^{\prime}|^2 +|C^{\prime}|^2) 
\Bigg[\lambda- \frac{\hat{u}(\hat{s})^2}{3} \Bigg]} \nn\\
&+& 4 |C^{\prime}|^2  \hat{m}_\ell^2 (2+2 \hat{m}_K^2-\hat{s})
+ 8 Re( C^{\prime} D^{\prime *}) \hat{m}_\ell^2 (1-\hat{m}_K^2)  
+4 |D^{\prime}|^2  \hat{m}_\ell^2 \hat{s} 
+ |X^{\prime} |^2 (2\hat{s}-4 \hat{m}_\ell^2) \nn\\
&+&4 Re(D^{\prime}{X^{\prime} }^*)  \hat{m}_\ell \hat{s} 
+4 Re(C^{\prime}{X^{\prime} }^*)\hat{m}_\ell (1-\hat{m}_K^2) 
\Bigg\}.
\label{eq:BKll}
\end{eqnarray}

In comparison to $B\to K^{(*)}\ell^+\ell^-$, theoretical calculations of the decay amplitudes for $B \to K^{(*)}\nu\bar\nu$ are considerably more reliable, owing
to the absence of long-distance interactions that affect
charged-lepton channels. We consider the missing energy distribution of the decay rates in terms the energy of the neutrino pair in the B rest frame and define the dimensionless
variable $\hat x = \hat E_{miss}$, which varies in the range $(1-\hat m_{K^{(*)}}^2)/2<\hat x < 1-\hat m_{K^{(*)}}$. For the $B \to K^{*}\nu\bar\nu$ channel, the partial rate then reads~\cite{Buchalla:2000sk,Colangelo:2006vm}
\begin{eqnarray}
\frac{d\Gamma}{d \hat x} &=& \frac{ G_F^2
   \alpha _{ \mathrm{em}}^2 m_B^5}{128 \pi ^5} \left| V_{t,b}
   V_{t,s}^*\right|^2  
   \left|\frac{ {\delta
   C}_{ \nu\bar\nu}+ {X}_{ {t}}}{\sin^2 \theta_W}\right|^2 \lambda^{1/2} \left\{\hat{s}
   \left[\left(\hat{m}_{K^*}+1\right)
   A_1\left(\hat{s}\right)-\frac{\lambda^{1/2}
   V\left(\hat{s}\right)}{\hat{m}_{K^*}+1}\right]^2\right.\nn\\
&& \left.+\hat{s} \left[ \left(\hat{m}_{K^*}+1\right)
   A_1\left(\hat{s}\right)+\frac{\lambda^{1/2}
   V\left(\hat{s}\right)}{\hat{m}_{K^*}+1}\right]^2\right.\nn\\
&&\left. +\left[\frac{\left(\hat{m}_{K^*}+1\right)^2
   \left(\sqrt{\lambda+4\hat{m}_{K^*}^2}-2\hat{m}_{K^*}^2\right
   ) A_1\left(\hat{s}\right)-\lambda
   A_2\left(\hat{s}\right)}{2\hat{m}_{K^*}
   \left(\hat{m}_{K^*}+1\right)}\right]^2\right\}\,,
\end{eqnarray}
while in the case of  $B\to K\nu\bar\nu$ we have
\begin{equation}
\frac{d\Gamma}{d \hat x} = \frac{G_F^2 \alpha _{\mathrm{em}}^2 m_B^5}{128
   \pi ^5 } \left|  V_{t,b}
   V_{t,s}^*\right|^2
   \left|\frac{{\delta
   C}_{{\nu\bar\nu}}+{X}_{{t}}}{\sin^2 \theta_W}\right|^2
   {f_+}(\hat s) {\lambda
   }^{3/2}\,.
\end{equation}

\subsection{$B\to X_s \gamma$}
Since the scalar-density operator does not contribute 
to $B\to X_s \gamma$, in this case the treatment is completely 
equivalent to the SM. We can thus take advantage of the 
complete NNLO analysis of Ref.~\cite{Misiak:2006zs}.
Expressing the branching ratio in terms of the initial 
conditions of $C_7$ and expanding around the SM value 
we can write~\cite{Misiak:2006zs,Misiak:2006ab,Misiak:private}
\begin{eqnarray}
\cB(B\to X_s \gamma)_{[E_{\gamma}>1.6~\mathrm{GeV}]} &=&
\left|\frac{V_{ts}V^*_{tb}}{V_{cb}}\right|^2\frac{6.00\times
10^{-5}}{C_{uc}} \left[3.15 - 7.18 \delta C_7 +4.74 (\delta C_7)^2 \right]\, .
\label{eq:bsg}
\end{eqnarray}
where $C_{uc} = |V_{ub}/V_{cb}|^2 \Gamma(B \to X_c e\nu)/ 
\Gamma(B \to X_u e\nu)$ and its numerical value
is reported in Table~\ref{table:inputs}. 
Following  Ref.~\cite{Misiak:2006zs} we assign a theoretical error 
of $0.23\times 10^{-4}$
to this expression.

Thanks to the precise experimental measurement of the $B\to X_s\gamma$
rate, Eq.~(\ref{eq:bsg}) 
provide a very stringent bound on non-standard 
contributions to electric-dipole and chromomagnetic operators. 
As anticipated, we do not treat $\delta C_8$ as an independent 
parameter in the fit: the bounds on $\delta C_7$ 
derived by means of Eq.~(\ref{eq:bsg}) should be interpreted
as bounds on the linear $\delta C_7$--$\delta C_8$
combination in Eq.~(\ref{eq:C7C8rel}).

\subsection{$B_{s,d} \to \ell^+ \ell^-$}
The pure leptonic decays $B_{s} \to \ell^+ \ell^-$
receive contributions only from the effective operators $P_{10}$ 
and $P^\ell_{0}$. These are free from the contamination of 
four-quark operators, which makes the generalization to the  
$b \to d$ case straightforward. 
 
The $B_{s} \to \ell^+ \ell^-$ rates can be written as 
\begin{equation}
\Gamma(B_s\to \ell^+ \ell^-) = \frac{\alpha_{\rm em}^2 G_F^2 |V_{tb}^*V_{ts}
A_{10}|^2}{16 \pi^3} m_{B_s} m_{\ell}^2 f_{B_s}^2
\sqrt{1-4\frac{ m^2_{\ell}}{m^2_{B_s} }} \left[|1+\delta_S|^2 + \left(1-4
\frac{ m^2_{\ell}}{m^2_{B_s} } \right)|\delta_S|^2\right]~,
\label{eq:Bsll}
\end{equation}
where 
\beq
\delta_S ~=~ \frac{C_{0}^\ell m_{B_s}^2}{ 4 A_{10} (m_b + m_s) m_{\ell} }
 ~=~   \frac{C_{0}^\ell }{\lambda_\ell \lambda_b } 
  \frac{ m_{B_s}^2 \tan\beta^2 }{  4 A_{10} \langle H_U \rangle^2 }
\frac{m_b}{m_b + m_s}~.
\eeq
The $B_{d} \to \ell^+ \ell^-$ rates are obtained 
from Eq.~(\ref{eq:Bsll}) with the exchange 
$\{V_{ts}$, $m_{B_s}$, $m_s\} \to \{V_{td}$, $m_{B_d}$, $m_d\}$.
Neglecting tiny corrections of $O(m_s/m_b)$, 
this leads to one of the most stringent tests of the 
MFV scenario, both at small and large $\tan\beta$ values, 
namely the relation
\beq
\frac{ \Gamma(B_s\to \ell^+ \ell^-) }{ \Gamma(B_d\to \ell^+ \ell^-) }
\approx \frac{  f_{B_s}m_{B_s} }{ f_{B_d}m_{B_d} }
\left|  \frac{ V_{ts} }{ V_{t_d} } \right|^2~.
\label{eq:Bll_p}
\eeq
On the other hand, we stress that the relation between 
$B_{s,d} \to \ell^+ \ell^-$ rates and $\Delta F = 2$ 
observables discussed in \cite{Buras:2003td} holds only in 
specific (constrained) versions of the MFV 
scenario.

\subsection{$K  \to \pi \nu \bar \nu $}

The branching ratio of charged and neutral 
$K \to \pi \nu \bar \nu $ decays can be simply expressed as
\beq
\mathcal{B}\left(  K^{+}\rightarrow\pi^{+}\nu\bar{\nu} \right)   
 =\kappa_{\nu}^{+}\left(1+\Delta_{EM}\right)  \displaystyle |y_{\nu}|^2~, \qquad 
\mathcal{B}\left(  K_L \rightarrow\pi^{0}\nu\bar{\nu} \right)   
 =\kappa_{\nu}^{L}  \displaystyle [\im(y_{\nu})]^2~,
\eeq
where from~\cite{Antonelli:2008jg,Mescia:2007kn} $
\kappa_{\nu}^+=0.7867(43) \times 10^{-5}$ 
and $\kappa_{\nu}^{L}=3.3624(264)\times 10^{-5}$ whereas   
the electromagnetic corrections $\Delta_{EM}=-(0.30 \pm 0.05)\%$\,.
Similarly to $B_{s,d} \to \ell^+ \ell^-$ rates, also in this case we have 
two observables controlled by a single free parameter: the real 
coefficient $\delta C_{\nu\bar\nu}$ in Eq.~(\ref{NUCoeff}). 
Thus also in this case the ratio of the  two $K \to \pi \nu \bar \nu $ rates 
leads to a significant model-independent test of the MFV hypothesis.

\begin{table}
\begin{center}
\begin{tabular}{ll}
  Quantity & Value \\
  \hline
  \hline
  $|V_{us}|$ & $0.2255\pm 0.0007$~\cite{Antonelli:2008jg} \\
  $|V_{cb}|$ & $(4.1\pm 0.1)\times 10^{-2}$ \cite{HFAG}   \\
  $|V_{ub}|$ & $(3.8\pm 0.4)\times 10^{-3}$ \cite{HFAG}  \\
  $\gamma$ & $(70.3 \pm 6.3)^{o}$ \cite{UTfit} \\
  \hline
  $f_{B_s}$ & $(0.260\pm 0.030)~\mathrm{GeV}$ \cite{fbslattice}\\
  $f_{B_s}/f_{B_d}$ & $(1.21\pm 0.06)$  \cite{Tantalo} \\
  \hline
  $\cB(B \to X_c \ell \nu)$ & $(10.75\pm 0.16)\times 10^{-2}$ \cite{HFAG} \\
  $C_{uc}$ & $0.58 \pm 0.01$ \cite{Misiak:2006zs}  \\
\end{tabular}
\caption{\label{table:inputs} 
Main inputs used in the numerical analysis (inputs not explicitly 
indicated in this Table are taken from Ref.~\cite{PDG08}).}
\end{center}
\end{table}

\section{Numerical analysis}
\label{sect:bounds}

In order to determine the presently allowed range 
of the $\delta C_i$ we have performed a global
fit of the $\Delta F=1$ observables in Table~\ref{tab:obs}.
The main numerical inputs beside rare processes
used in the fits are reported in Table~\ref{table:inputs}.
The latter have been assumed to be not correlated 
and not polluted by new physics 
effects. In particular,
as far the CKM angle $\gamma$ is concerned, we have used 
the results of Ref.~\cite{UTfit} where the CKM matrix 
is determined using only tree-level observables.
Our fitting procedure follows the method adopted by 
the UTFit collaboration \cite{UTfit}: we integrate over the 
probability distributions of inputs and conditional probability distributions of observables assuming validity of MFV to obtain (after proper normalization) the probability distributions for 
the $\delta C_i$.

\begin{table}[t]
\begin{center}
\begin{tabular}{lll}
  $\delta C_i$ & $95\%$ probability bound& Observables \\
  \hline\hline
  $\delta C_7$ & $[-0.14,0.06]\cup [1.42,1.62]$ & $B\to X_s \gamma$, $B\to X_s \ell^+ \ell^-$\\
  $\delta C_9$ & $[-2.8,0.8]$ & $B\to X_s \ell^+ \ell^-$\\
  $\delta C_{10}$ & $[-0.4,2.3]$ & $B\to X_s \ell^+ \ell^- $, $B_s\to \mu^+ \mu^-$\\
  $\delta C^\mu_{S}/m_b$  &   $[-0.09,0.09]/({\rm 4.2 GeV})$ &  $B_s\to  \mu^+ \mu^-$\\  
  $\delta C_{\nu\bar\nu}$ & $[-6.1,2.0]$ & $K^+\to \pi^+ \nu \bar \nu$
\end{tabular}
\caption{\label{table:bounds} 
Combined bounds on the effective Wilson coefficients
   in the MFV scenario, and observables used to derive the bounds
   (in the case of the scalar operator we report the bound in
   terms of the
   scale-independent combination $\delta C^\mu_{S}/m_b$).}
\end{center}
\end{table}

\begin{table}[t]
\begin{minipage}{\textwidth}
\begin{center}
\begin{tabular}{lll}
  Operator & $\Lambda_i@95\%$~prob.~[TeV] & Observables \\
  \hline\hline
$H_D^\dagger \left( {\bar D}_R  \lambda_d 
\lambda_{\rm FC} \sigma_{\mu\nu} Q_L \right) (e F_{\mu\nu})$
& $6.1$ & $B\to X_s \gamma$, $B\to X_s \ell^+ \ell^-$\\

$H_D^\dagger \left( {\bar D}_R  \lambda_d
\lambda_{\rm FC} \sigma_{\mu\nu}  T^a  Q_L \right) (g_s G^a_{\mu\nu})$
& $3.4$ & $B\to X_s \gamma$, $B\to X_s \ell^+ \ell^-$\\

$\left( {\bar Q}_L \lambda_{\rm FC} \gamma_\mu Q_L
\right) (e D_\mu
F_{\mu\nu})$

& $1.5$ &  $B\to X_s \ell^+ \ell^-$\\

$ i \left( {\bar Q}_L \lambda_{\rm FC} \gamma_\mu
Q_L \right) H_U^\dagger D_\mu H_U$
& $1.1^a$ & $B\to X_s \ell^+ \ell^-$, $B_s\to\mu^+\mu^-$\\
$ i \left( {\bar Q}_L \lambda_{\rm FC}\tau^a \gamma_\mu
Q_L \right) H_U^\dagger\tau^a  D_\mu H_U$
& $1.1$\footnote{A discrete ambiguity is removed at $90\%$ probability, improving the bound to $2.3$~TeV.} & $B\to X_s \ell^+ \ell^-$, $B_s\to\mu^+\mu^-$\\
$\left( {\bar Q}_L \lambda_{\rm FC} \gamma_\mu Q_L \right) 
({\bar L}_L \gamma_\mu L_L)$
& $1.7$ &  $B\to X_s \ell^+ \ell^-$, $B_s\to\mu^+\mu^-$\\
$\left( {\bar Q}_L \lambda_{\rm FC} \gamma_\mu
\tau^a Q_L \right)({\bar L}_L \gamma_\mu  \tau^a L_L)$
& $1.7$ &  $B\to X_s \ell^+ \ell^-$, $B_s\to\mu^+\mu^-$\\
$\left( {\bar Q}_L \lambda_{\rm FC} \gamma_\mu
Q_L \right) ({\bar E}_R \gamma_\mu E_R)$  
& $2.7$ & $B\to X_s \ell^+ \ell^-$, $B_s\to\mu^+\mu^-$
\end{tabular}
\end{center}
\end{minipage}
\caption{\label{table:bounds2} Individual bounds on the scale of new physics 
for the most relevant MFV operators.}
\end{table}
 
\begin{figure}[t]
\begin{center}
\scalebox{0.5}{\includegraphics{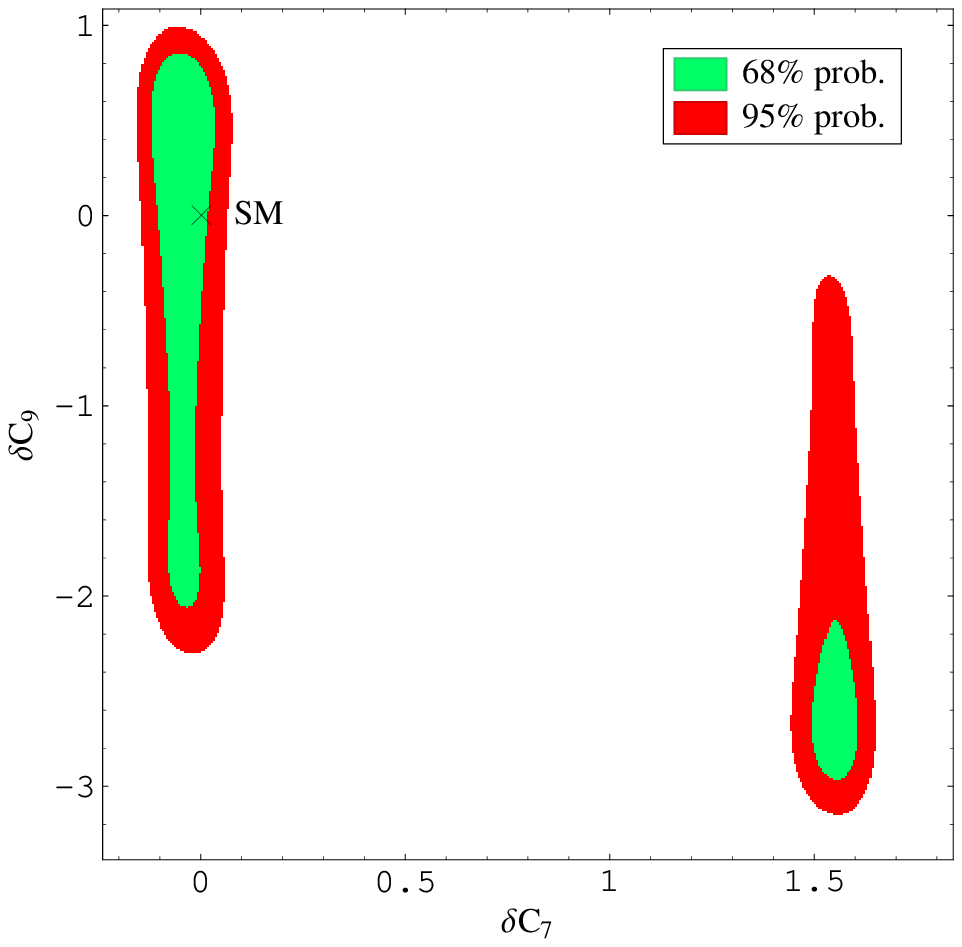}}
\scalebox{0.5}{\includegraphics{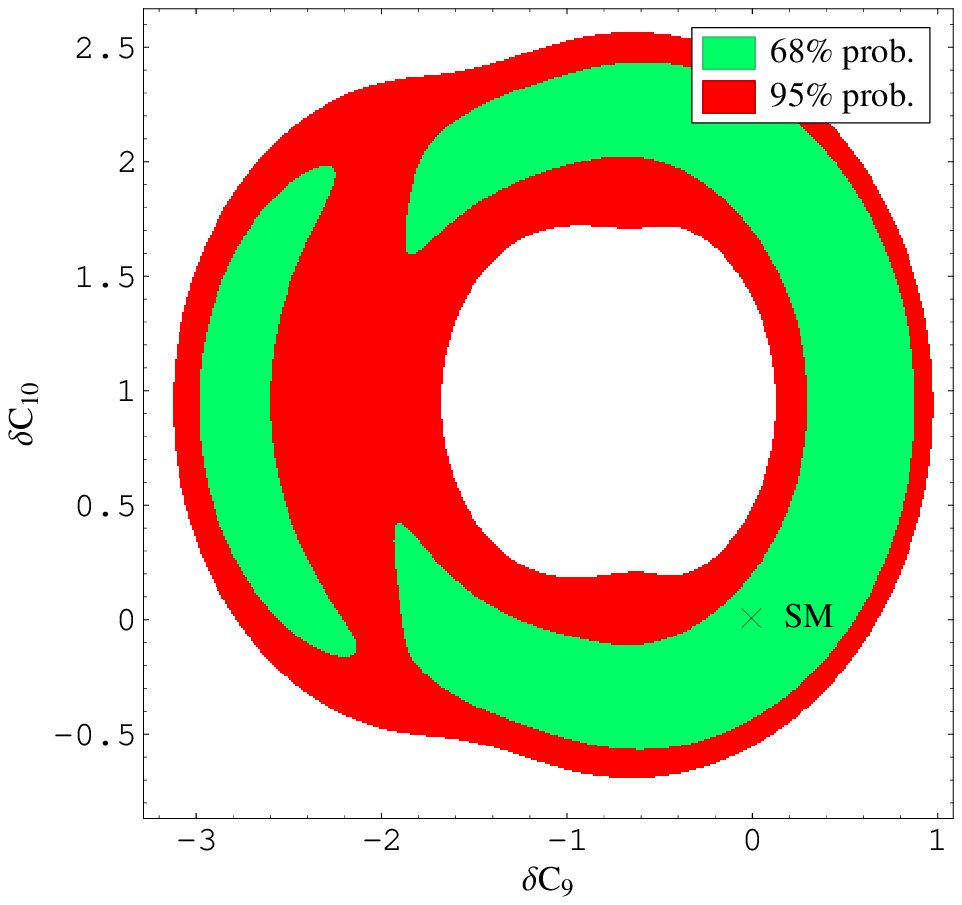}}
\scalebox{0.5}{\includegraphics{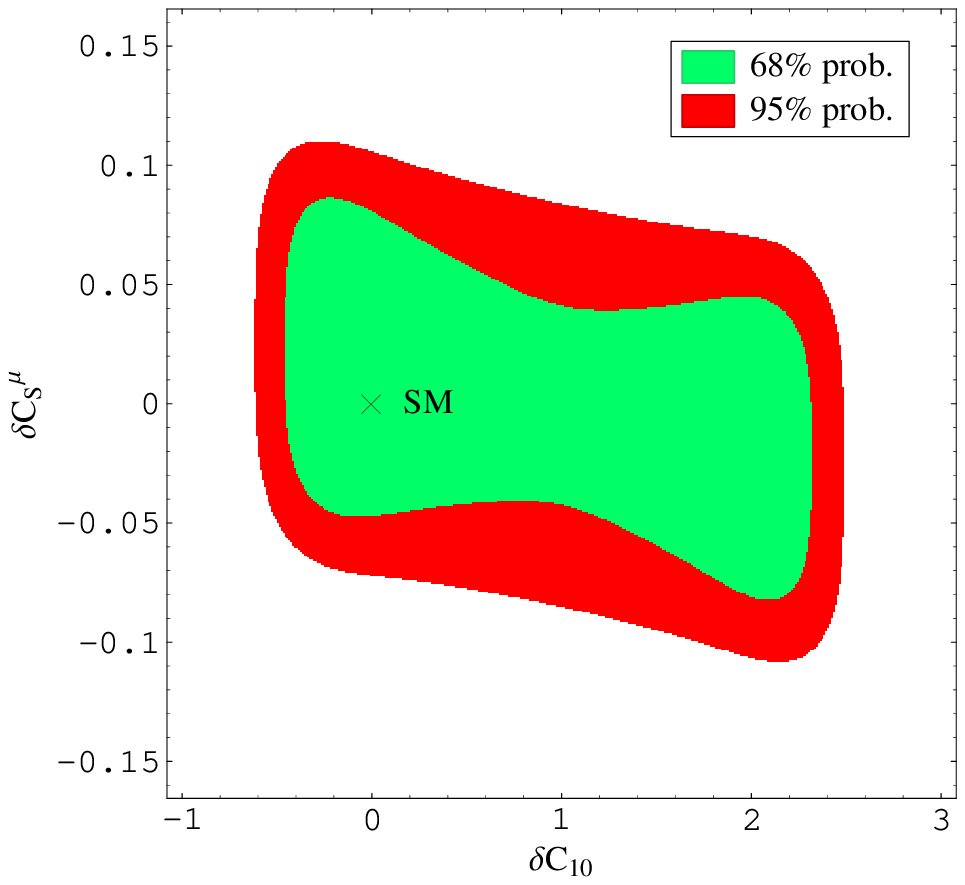}}
\caption{\label{fig:cor}Correlation plots for the pairs of shifts in the Wilson coefficients. Allowed regions at 68\% (in green) and 95\% (in red) probability.}
\end{center}
\end{figure}

The resulting ranges for the $\delta C_i$ and the corresponding 
bounds on the scale of new physics for the various operators
are shown in Tables~\ref{table:bounds} and~\ref{table:bounds2},
and in Fig.~\ref{fig:cor}. The bounds in Table~\ref{table:bounds}
are the results of the global fit, where all the $\delta C_i$
are allowed to vary. The most interesting correlations 
among pairs of $\delta C_i$ of the global fit 
are shown in  Fig.~\ref{fig:cor}. On the other hand, 
the bounds in Table~\ref{table:bounds2} 
correspond to the bound on the scale of each 
non-standard operator, assuming the others to
have a negligible impact. Note that the correlations 
of  the $\delta C_i$ play a non-trivial role 
also in Table~\ref{table:bounds2}, by means of 
Eq.~(\ref{eq:C9V}): each bound corresponds to setting one of the 
the $a_i=\pm1$ and the others to zero. 
In case of sign ambiguities, 
the bound on the scale corresponds 
to the lower allowed value. 

In the case of the scalar-density operator, the translation 
of the bound on $\delta C^\mu_{S}$
into a bound on the scale is not straightforward 
as for the other operators. Assuming that the 
coefficient of $\cO_{S1}$ in Eq.~(\ref{eq:basis1})
does not depend on $\tan\beta$ and 
setting $a_{S1}=\pm 1$ we get
\beq
\Lambda_{[O_{S1}]} > \Lambda_0 \left( \frac{ \lambda_b \lambda_\mu }{ 
\left| \delta C^\mu_{S} \right|_{\rm max} } \right)^{1/2} 
= (1.5~{\rm TeV}) \times \left( \frac{\tan\beta}{50} \right)~[95\%~{\rm prob.}]~.
\eeq
This bound, comparable to most of the bounds
in Table~\ref{table:bounds2}, is especially interesting in specific 
models, where it can be identified with a bound on the mass 
of heavy Higgs fields. This happens for instance in the MSSM, where 
$a_{S1}$ is suppressed by $1/16\pi^2$ ($\cO_{S1}$ being forbidden at the tree 
level by the Peccei-Quinn symmetry) but grows linearly 
with $\tan\beta$~\cite{D'Ambrosio:2002ex}.
In particular, setting 
$|a_{S1}|/\Lambda^2 = \tan\beta/(16\pi^2 M_H^2 )$, leads to 
\beq
M_H  >  (830~{\rm GeV}) \times \left( \frac{\tan\beta}{50} \right)^{3/2}~[95\%~{\rm prob.}]~.
\eeq

\begin{figure}[t]
\begin{center}
\scalebox{0.75}{\includegraphics{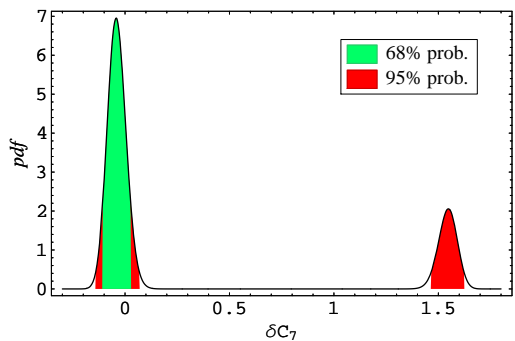}}
\caption{\label{fig:C7} Probability distribution of $\delta C_7$.}
\end{center}
\end{figure}

As far the sign of $C_7$ is concerned, we find that
the {\em wrong sign} solution to $C_7$ is still
allowed, but has a lower probability compared to the SM sign. 
As shown in Fig.~\ref{fig:C7}, the 
large contribution to $\delta C_7$,
corresponding to a sign flip of $C_7(m_b)$, 
has a probability of about 30\%. 
We stress that the sign flip of $C_7(m_b)$ occurs only 
if $C_9$ receives a sizable non-standard contribution. 
This is consistent with 
the conclusion of Ref.~\cite{Gambino:2004mv}, where 
the wrong sign solution to $C_7$ has been 
excluded assuming small new-physics effects in 
the other Wilson coefficients.

\begin{figure}[t]
\begin{center}
\scalebox{0.7}{\includegraphics{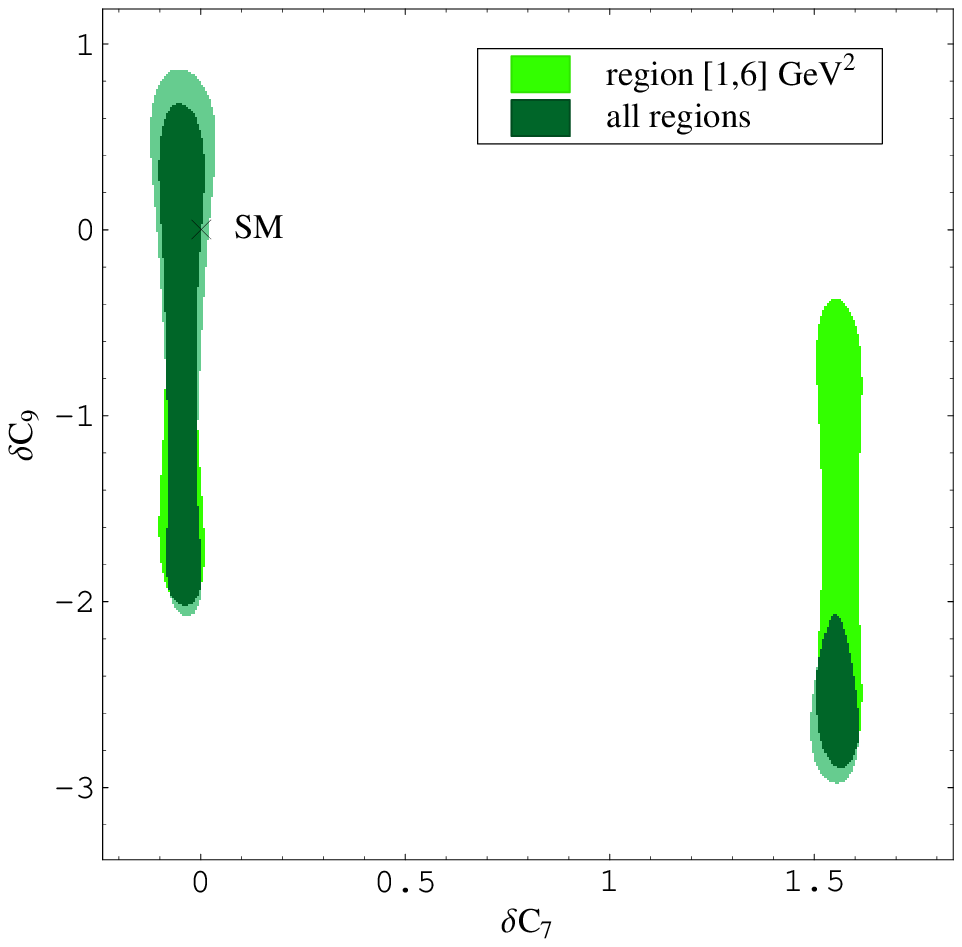}}
\scalebox{0.7}{\includegraphics{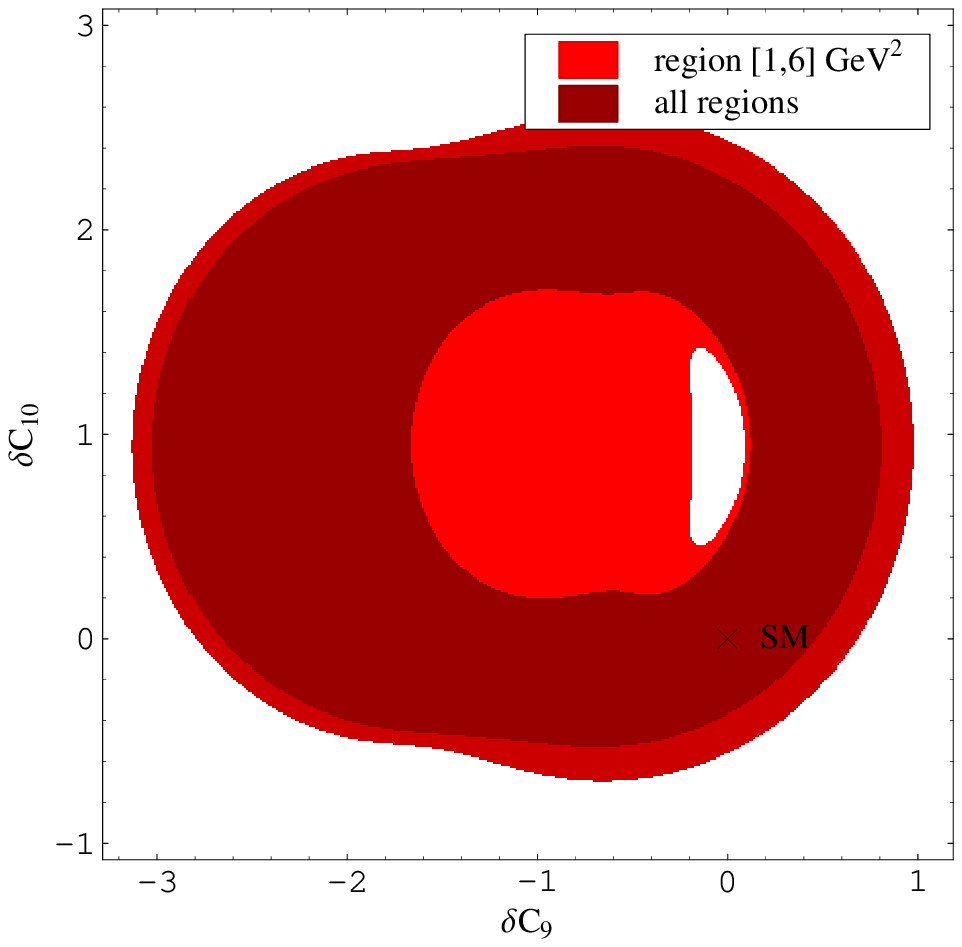}}
\caption{\label{fig:cor1a}Correlation plots for the pairs of shifts in the Wilson coefficients, which are most affected by including the low and high energy regions in $B\to X_s \ell^+\ell^-$. The left plot shows regions allowed at 68\% (in green) while in the right plot, the regions shown are allowed at 95\% (in red). The dark and light shaded regions correspond to the fit with and without the low and high energy regions in $B\to X_s \ell^+\ell^-$ respectively.}
\end{center}
\end{figure}

The impact of the low- and high-energy regions 
in $B\to X_s \ell^+\ell^-$, which are often neglected, 
can be seen in Fig.~\ref{fig:cor1a}, where we plot the most interesting 
68\% and 95\% allowed regions with or without the information
of these two measurements. In view of future 
experimental improvements, we report below 
the numerical values of the main observables 
expanded in powers of the $\delta C_i$:
\bea
&&  \cB(B \to X_s \gamma)(E_{\gamma}>1.6\mathrm{GeV}) = 3.13(23) \times
  10^{-4} \times(1
  -2.28\delta C_7
  +1.51\delta C_7^2)\,,
\eea
\bea
&& \cB(B \to X_s \ell^+ \ell^-)(q^2\in [0.0021,0.04]~\mathrm{GeV}^2) =
  0.8(2) \times 10^{-6} \nonumber\\
  &&\hspace{1cm}\times \left[1
  + 2.02 \delta C_{7}^2
  +0.20
   \delta C_{9}^2
   +0.20
   \delta C_{10}^2
   -2.23 \delta C_{7}
   +0.23
   \delta C_{9}
   -0.37 \delta C_{10}
   +0.31 \delta C_7 \delta C_9
 \right.\nonumber\\
&& \hspace{1cm} \left.
+0.01 (\delta C^\mu_{S})^2 - 0.01 \delta C_{10} \delta C^\mu_{S} 
+ 0.01 \delta C^\mu_{S} \right]\,\nn\\
 && \cB(B \to X_s \ell^+ \ell^-)(q^2\in [1.0,6.0]~\mathrm{GeV}^2) =
  1.6(1) \times 10^{-6} \nonumber\\
  &&\hspace{1cm}\times \left[ 1
  + 0.64 \delta C_{7}^2
  +0.61
   \delta C_{9}^2
   +0.61
   \delta C_{10}^2
   -0.03 \delta C_{7}
   +0.86
   \delta C_{9}
   -1.15 \delta C_{10} 
   +0.73 \delta C_7 \delta C_9\right.\nonumber\\
&& \hspace{1cm} \left. +0.13 (\delta C^\mu_{S})^2 - 0.03 \delta C_{10} 
\delta C^\mu_{S} + 0.04 \delta C^\mu_{S} \right]\,,\nonumber\\
&&  \cB(B \to X_s \ell^+ \ell^-)(q^2\in [14.4,25.0]~\mathrm{GeV}^2) =
  2.3(7) \times 10^{-7} \nonumber\\
  && \hspace{1cm}\times \left[ 1
  + 0.05 \delta C_7^2
  +0.28
   \delta C_7+0.59 \delta C_9^2
   +0.59
   \delta C_{10}^2
   +1.04 \delta C_9
   -1.13 \delta C_{10}
   +0.32 \delta C_7 \delta C_9 \right. \nonumber\\
&& \hspace{1cm} \left. +0.27 (\delta C^\mu_{S})^2 
- 0.02 \delta C_{10} \delta C^\mu_{S} 
+ 0.02 \delta C^\mu_{S} \right]\,,
\eea
\bea
 \cB(B_s \to \mu^+ \mu^-) &=& 4.1(8) \times 10^{-9}  \times \left[(1 - 1.04 \delta C_{10}
  - 29.3 \delta C^\mu_{S})^2
  + 860 (\delta C^\mu_{S})^2\right]\,,
\eea
\bea
   \cB(K^+ \to \pi^+ \nu \bar \nu ) &=& 8.6(9)\times 10^{-11} 
\left[ 1+ 0.96 \delta C_{\nu\bar\nu} 
+ 0.24\delta C_{\nu\bar\nu}^2\right]\,.
\eea

Since the experimental information of exclusive $B\to K^*\ell^+\ell^-$ 
modes is quite good~\cite{babar:2008ju}, we have also investigated 
the impact of including these observables in the fit. 
In this case we have used the results of Ref.~\cite{Ball:2004rg}
for the hadronic form factors and their corresponding errors, taking into account additional theoretical uncertainty due to neglected additional long distance effects~\cite{Beneke:2001at,Beneke:2004dp,Ali:2006ew}.

\begin{figure}[t]
\begin{center}
\scalebox{0.7}{\includegraphics{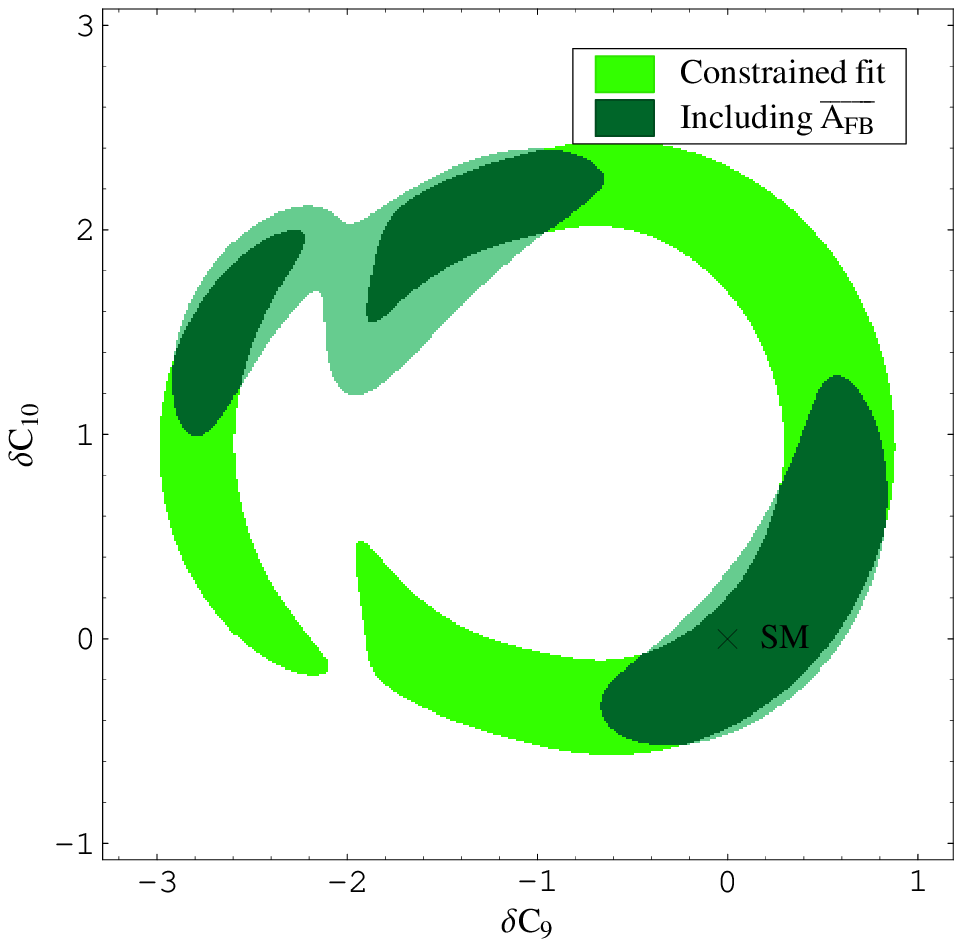}}
\scalebox{0.7}{\includegraphics{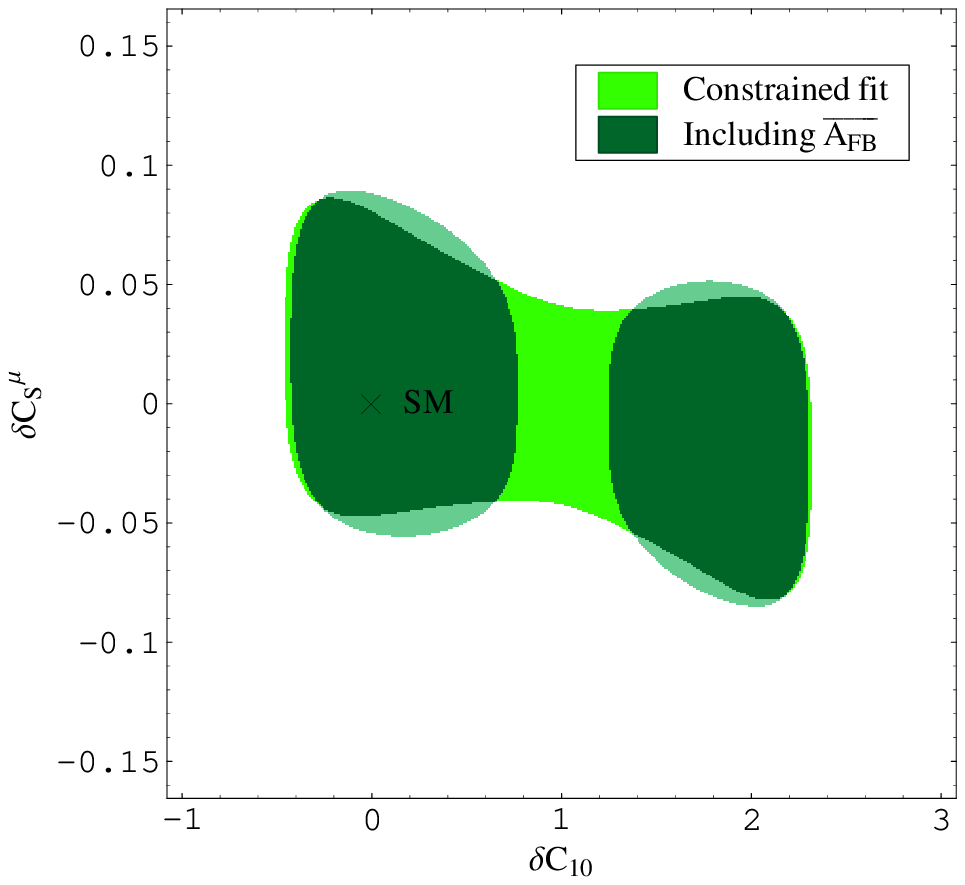}}
\caption{\label{fig:cor2}Correlation plots for the pairs of shifts in the Wilson coefficients, which are most affected by including the exclusive $A_{FB}$ observables. Allowed regions at 68\% with (dark green) and without (light green) exclusive $A_{FB}$ observables are shown.}
\end{center}
\end{figure}

The FB asymmetry plays a significant role. In particular, the normalised
FB asymmetry 
\beq
 \bar {\cal A}_{\rm FB}(q^2) = \frac{1}{\d \Gamma /\d q^2  }
 \frac{\d \a_{\rm FB}}{\d q^2}
\eeq
is interesting since the main uncertainty due to the  
form factors, namely the overall normalization of the decay rate,
partly cancels out. Moreover, $A_{FB}(B\to K^*\ell^+\ell^-)$ in the low 
$q^2$ energy region is very small in the SM due to the destructive 
interference of $C_7$ and $C_9$ (resulting in the famous zero
of the asymmetry at low $q^2$). It is therefore a good probe 
of the relative sign of the Wilson coefficients. 
The impact of including the presently available data on 
$\a_{\rm FB}(B\to K^*\ell^+\ell^-)$ is shown in 
Fig.~\ref{fig:cor2}.\footnote{$A_{FB}(B\to K^*\ell^+\ell^-)$
has been measured both by Belle~\cite{Ishikawa:2006fh} 
and Babar~\cite{babar:2008ju}. However, in Ref.~\cite{Ishikawa:2006fh}
only the fully integrated asymmetry has been reported. 
The normalised values of 
$A_{FB}(B\to K^*\ell^+\ell^-)$ in different $q^2$ bins,
as reported in Ref.~\cite{babar:2008ju}, 
represent the most useful information
for our purpose. For this reason, 
we have restricted our numerical analysis
only to the results in Ref.~\cite{babar:2008ju}.} 
As can be seen, at present the additional information significantly reduces part of the ambiguities 
in the $\delta C_9$--$\delta C_{10}$ and $\delta C_{10}$--$\delta C^{\mu}_0$
planes only at $68\%$ probability level. On the other hand, the overall 
bounds on the scales of individual coefficients 
do not change in appreciable way. This may seem at odds with conclusions reached in refs.~\cite{babar:2008ju,Ishikawa:2006fh}. However, the central experimental values for the high $q^2$ region lie approximately 1.5 standard deviations above the range of possible theory predicitions within MFV satisfying other bounds. In addition, this range of theory predictions spans less than two experimental standard deviations. Under the assumption of MFV validity, the present FB asymmetry measurements therefore cannot significantly affect the $95\%$ probability regions of $\delta C_i$. As seen on Fig.~\ref{fig:cor2}, the situation would however improve dramatically, once the experimental precision would approximately double.

Contrary to the FB asymmetry, it turns out that the
$K^*$ longitudinal polarization is not very sensitive 
to new physics in the MFV scenario. 
%

For completeness, we report below the numerical expressions of $A_{FB}(B\to K^*\ell^+\ell^-)$,
integrated and normalised to the decay rate 
as in Ref.~\cite{babar:2008ju}  
expanded in powers of the $\delta C_i$:
\bea
&&  
\frac{ \cA_{\rm FB}(B \to K^* \ell^+ \ell^-)_{q^2<6.25~\mathrm{GeV}^2} }{ \Gamma (B \to K^* \ell^+ \ell^-)_{q^2<6.25~\mathrm{GeV}^2} }
=  -0.01(2) \times \nonumber\\
  && \hspace{1cm}\times \frac{(1 - 20.   \delta C_7 - 11. \delta C_9   + 1. \delta C_{10} + 21. \delta C_7 \delta C_{10} + 11. \delta C_9 \delta C_{10} )}{(1  + 1.2 \delta C_7^2  -0.7   \delta C_7 +0.5 \delta C_9^2   +0.5   \delta C_{10}^2   +0.6 \delta C_9   -1.0 \delta C_{10}   +0.8 \delta C_7 \delta C_9 )} \, , \no \\
&& \frac{ \cA_{\rm FB}(B \to K^* \ell^+ \ell^-)_{q^2>10.24~\mathrm{GeV}^2}  }{ \Gamma(B \to K^* \ell^+ \ell^-)_{q^2>10.24~\mathrm{GeV}^2} }
=  0.20(8) \times \nonumber\\
  && \hspace{1cm}\times \frac{(1 + 0.5   \delta C_7 + 1.2 \delta C_9   -1.0 \delta C_{10} - 0.5 \delta C_7 \delta C_{10} - 1.3 \delta C_9 \delta C_{10} )}{(1  + 0.1 \delta C_7^2  +0.4   \delta C_7 +0.6 \delta C_9^2   +0.6   \delta C_{10}^2   +1.0 \delta C_9   -1.2 \delta C_{10}   +0.4 \delta C_7 \delta C_9 )} \, .
\qquad 
\eea
In the above expressions, we have neglected the scalar operator contributions, which are made negligible by the strong bound coming from $B_s\to\mu^+\mu^-$. However, $\delta C^{\mu}_S$ dominates possible NP effects in the lepton universality ratios of $B\to K^{(*)}\ell^+\ell^-$, as we discuss in the next section.
 

\section{Predictions}
\label{sect:pred}

\begin{figure}[t]
\begin{center}
\scalebox{0.7}{\includegraphics{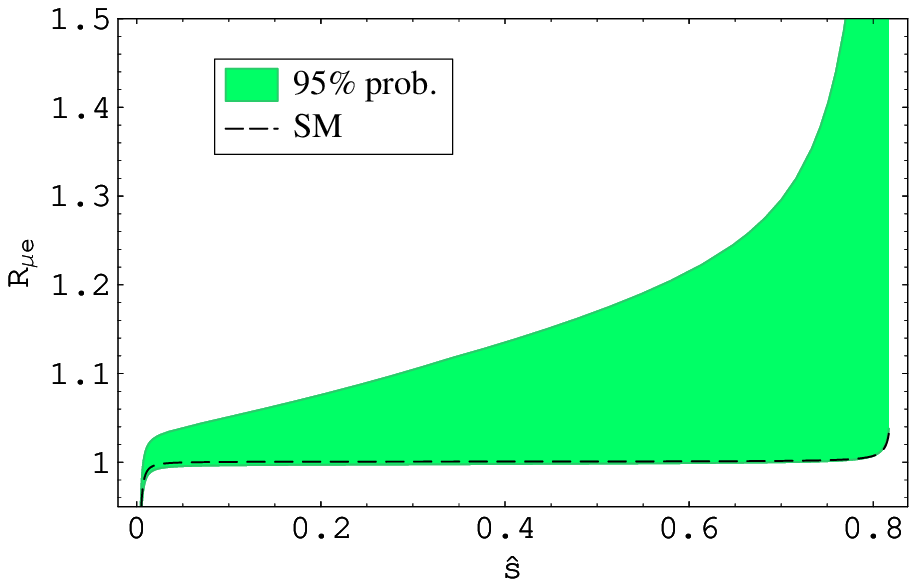}}
\scalebox{0.7}{\includegraphics{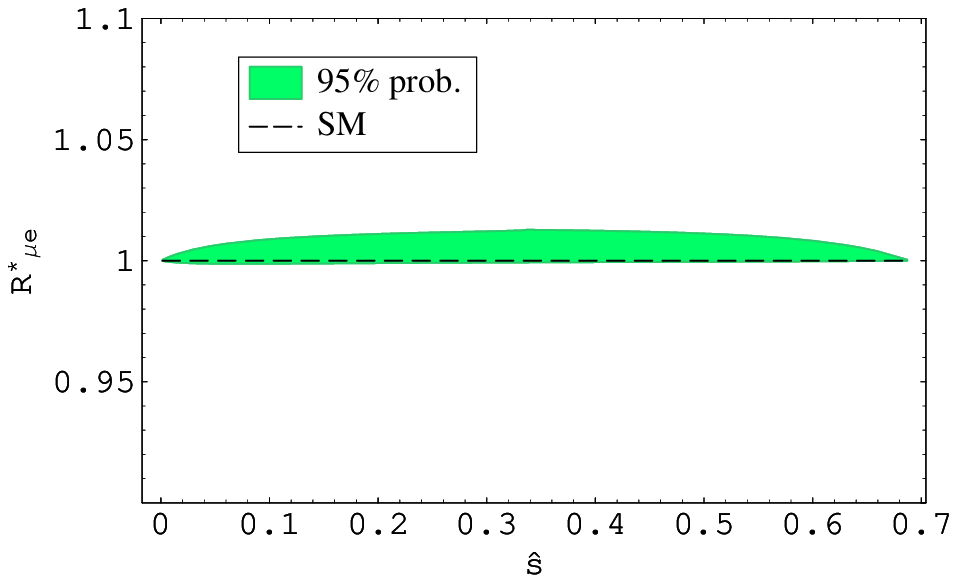}}
\caption{\label{fig:rmue}95\% probability allowed ranges 
for the lepton universality ratio 
(muon/electron) for the differential distributions of  $B\to K\ell^+\ell^-$ (left) 
and $B\to K^*\ell^+\ell^-$ (right).} 
\end{center}
\end{figure}

Using the bounds on the MFV operators discussed in  
the previous section, we obtain a series of constraints
for FCNC processes which are not well measured yet. 
The most interesting predictions can be summarised
as follows:

\begin{figure}[t]
\begin{center}
\scalebox{0.95}{\includegraphics{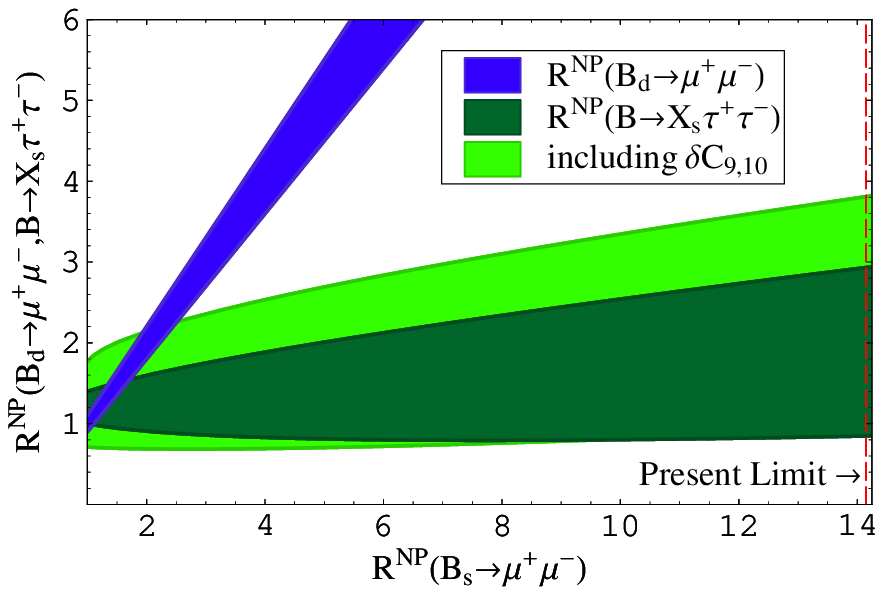}}
\caption{\label{fig:bsmuta}
Correlations between $B_s\to \mu^+\mu^-$,  
$B_d\to \mu^+\mu^-$, and $B_s \to X_s \tau^+\tau^-$.
For a given value of $\cB(B_s\to \mu^+\mu^-)$, 
normalised to the SM prediction (horizonatal axis), we plot the 
range of $\cB(B_d\to \mu^+\mu^-)$ and $\cB(B_s \to X_s \mu^+\mu^-)$,
both normalised to the correponding SM predictions (vertical axis).  }
\end{center}
\end{figure}

\begin{figure}[t]
\begin{center}
\scalebox{0.7}{\includegraphics{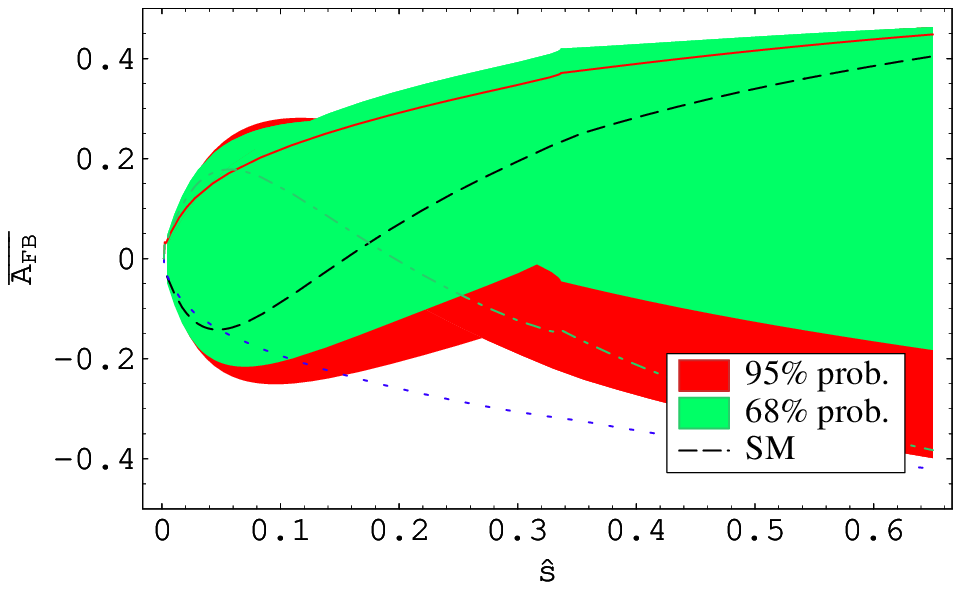}}
\scalebox{0.7}{\includegraphics{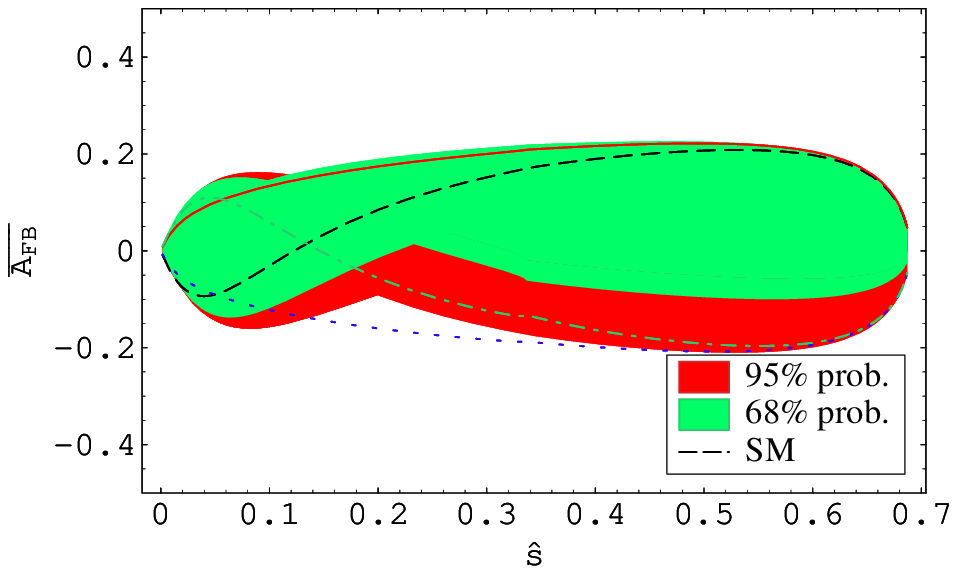}}
\caption{\label{fig:afb}
95\% probability allowed range for the inclusive normalised 
FB asymmetry in $B\to X_s \ell^+\ell^-$ (left) and $B\to K^*\ell^+\ell^-$ (right).
The dashed curve denotes the central value of the SM prediction. The other
three lines indicate the central values obtained flipping 
sgn($C_7/C_{10}$) and/or sgn($C_9/C_{10}$). }
\end{center}
\end{figure}

\begin{itemize}
\item{$R^{(\mu/e)}(B \to K^{(*)} \ell^+ \ell^-)$}. \\
As pointed out in~Ref.~\cite{Hiller:2003js} the 
lepton universality  ratios (or the muon to electron ratios 
of branching ratios) in $B \to K^{(*)} \ell^+ \ell^-$ 
are very clean probes of possible scalar density operators. 
Within the MFV framework there is a one-to-one correspondence 
between possible deviations from the SM in the 
lepton universality ratios and the contribution of  
the scalar-density operator in $B_s \to \mu^+\mu^-$. 
Thanks to the substantial improvement on $B_s \to \mu^+\mu^-$,
the effect is highly constrained. 

The present allowed ranges for the differential distributions 
of the lepton universality ratios are shown in Fig.~\ref{fig:rmue}.
As can be seen, the effect is negligibly small in the 
$B\to K^*$ case. In the $B\to K$ case there is still some room 
for deviations from the SM, but only in the high-$q^2$ region, 
where the decay is suppressed. The maximal integrated 
effect is reported in Table~\ref{tab:obs2}.
Note the major improvement with respect to 
the analysis of Ref.~\cite{Hiller:2003js}, where the
loose bound on $B_s \to \mu^+\mu^-$ allowed much larger
non-standard effects. 

\item{$\cB(B_d \to \mu^+ \mu^-)$}\\
As anticipated, probably the best test of the MFV scenario 
in the $\Delta F=1$ sector is the ratio in Eq.~(\ref{eq:Bll_p}).
Taking into account the present bound on $B_s\to \mu^+\mu^-$ 
this implies the upper limit
\beq
\cB(B_d \to \mu^+ \mu^-)<1.2 \times 10^{-9}~,
\eeq
at 95\% probability. The clear correlation between $\cB(B_d \to \mu^+ \mu^-)$
and $\cB(B_s \to \mu^+ \mu^-)$ in the  MFV framework 
is illustrated in Fig.~\ref{fig:bsmuta}.

\item{$\cB(B \to X_s \tau^+ \tau^-)$} \\
The bound on the scalar-current operator allows us to derive 
a non-trivial bound also on $\cB(B \to X_s \tau^+ \tau^-)$, 
which at large $\tan\beta$ is sensitive to 
scalar-current operators. 
The parametrical dependence on the modified Wilson 
coefficient of the integrated branching ratio is 
\begin{eqnarray}
  \cB(B \to X_s \tau^+ \tau^-)_{q^2\in [14.4,25.0]~\mathrm{GeV}^2} &=&
  1.6(5) \times 10^{-7} \nonumber\\
  && \hspace{-6cm}\times (1
  + 0.06 \delta C_7^2
  +0.38
   \delta C_7+0.81 \delta C_9^2
   +0.41
   \delta C_{10}^2
   +1.41 \delta C_9 \nonumber\\
&&\hspace{-6cm}
   -0.78 \delta C_{10}
   +0.43 \delta C_7 \delta C_9 - 0.54 \delta C_{10}\delta C^{\tau}_{S} 
  + 0.52 \delta C^{\tau}_{S} +0.32 {\delta C^{\tau}_{S}}^2 )\,.
\end{eqnarray}
Assuming $\delta C^{\tau}_{S} =  (m_{\tau}/m_{\mu})\delta C^{\mu}_{S} $,
in agreement with Eq.~(\ref{eq:C9V}), 
and taking into account the allowed ranges of the Wilson 
coefficients in Table~~\ref{table:bounds}, we obtain 
the $95\%$~probability bound reported in Table~\ref{tab:obs2}.

\item{$\mathrm{d}A_{FB}/\mathrm{d}q^2 (B \to X_s \ell^+ \ell^-)$} \\
Taking into account the available constraints on $C_{7,9,10}$, the 
resulting allowed range for the inclusive FB asymmetry is shown in 
Fig.~\ref{fig:afb}. As can be seen, in this case there is still
a large room for non-standard effects: this observable is one 
of the few examples of quantities which could exhibit  large 
deviations from the SM even in the pessimistic MFV framework. 
On the other hand, present data exclude most confgurations with flipped sgn($C_9/C_{10}$) at $68\%$ probability.

\item{$\cB(B\to K^{(*)} \nu \bar \nu)$ and $\cB(K_L\to\pi^0\nu\bar\nu)$} \\
From the $K^+\to\nu\bar \nu$ bound on $\delta C_{\nu\nu}$ we can
bound the rates of all the $\nu \bar \nu$-type FCNC 
transitions\footnote{These predictions are valid only in the limit where we can neglect operators with the $Y_u^{\dagger}Y_u Y_d^{\dagger}Y_d$ flavour structure. This condition is always fullfiled at small/moderate $\tan\beta$ and even at large $\tan\beta$ holds in most excplicit MFV scenarios.}. The most
interesting predictions are reported in Table~\ref{tab:obs2}
(in the $\cB(B\to K^{(*)} \nu \bar \nu)$ case we take into 
account form-factor uncertainties along the lines discussed 
in Section~\ref{sect:exclusive}).
For completeness, we give the numerical expressions of the  
$K_L\to \pi^0 \nu\bar \nu$ and $B\to K^{(*)} \nu \bar \nu$ branching ratios:
\bea
   \cB(K_L \to \pi^0 \nu \bar \nu )& =& 2.9(5)\times 10^{-11}
\left[1+ 1.37 \delta C_{\nu\bar\nu} 
+ 0.47\delta C_{\nu\bar\nu}^2\right]\,,\nn\\
   \cB(B\to K^{} \nu \bar \nu)& =& 0.5(1)\times 10^{-5}
\left[1+ 1.2 \delta C_{\nu\bar\nu} 
+ 0.4\delta C_{\nu\bar\nu}^2\right]\,,\nn\\
   \cB(B\to K^{*} \nu \bar \nu)& =& 1.2(3)\times 10^{-5}
\left[1+ 1.2 \delta C_{\nu\bar\nu} 
+ 0.4\delta C_{\nu\bar\nu}^2\right]\,.
\eea

\end{itemize}

\begin{table}[t]
\begin{minipage}{\textwidth}
\begin{center}
\begin{tabular}{lllll}
 Observable & Experiment  & {MFV bound} & {SM prediction} \\
  \hline
  \hline
  $R^{(\mu/e)}(B \to K \ell^+ \ell^-)-1$ & $0.17\pm 0.28$~\cite{Abe:2004ir,Aubert:2008ps}\footnote{Here we quote na\"ive averages of the values obtained by the experiments and with symmetrized errors.} & $[-0.004,0.14]$ & $O (10^{-4})$~\cite{Bobeth:2007dw}  \\
  \hline
  $R^{(\mu/e)}(B \to K^* \ell^+ \ell^-)-1$ & $0.37^{+0.53}_{-0.40}\pm 0.09$~\cite{Aubert:2008ps}
  & $[-0.002,0.01]$ & $\lsim 10^{-2}$ \\
  \hline
  $\cB(B_d \to \mu^+ \mu^-)$ & $<1.8 \times 10^{-8}$~\cite{:2007kv}
  \qquad    & $<1.2 \times 10^{-9}$ & $1.3(3)\times 10^{-10}$ \\
  \hline
  $\cB(B \to X_s \tau^+ \tau^-)$ & -- & 
  $ < 5 \times 10^{-7}$ & $1.6(5)\times 10^{-7}$ \\
  \hline
  $\cB(B\to K\nu\bar\nu)$   & \cite{:2007zk} &  $<0.4\times 10^{-4}$
& $(0.5 \pm 0.1)\times 10^{-5}$ \\
 \hline
  $\cB(B\to K^*\nu\bar\nu)$ & \cite{:2007zk} &  $<9.4\times 10^{-5}$
& $(1.2 \pm 0.3)\times 10^{-5}$ \\
  \hline
  $\cB(K_L \to \pi^0 \nu \bar \nu)$ & \cite{Nix:2007sa} & $<2.9\times 10^{-10}$ & $2.9(5)\times 10^{-11}$ \\
\end{tabular}
\end{center}
\end{minipage}
\caption{Predicted observables. All bounds are 95\% probability limits.
In the first two lines, the SM predictions refer to the kinematical
region where lepton-mass phase-space effects can be 
safely neglected. The experimental results on $\nu\bar\nu$ modes 
are not explicilty indicated 
since  only 90\% C.L.~limits are available. \label{tab:obs2} }
\end{table}

\section{Conclusions}

The MFV hypothesis provides an efficient  
tool to analyse flavour-violating 
effects in extensions of the SM. 
The effective theory built on this 
symmetry and  and symmetry-breaking hypothesis
leads to: 1) a natural suppression of flavour violating 
effects, in agreement with present observations,
even for for new physics in the TeV range;
2) a series of experimentally testable 
predictions, which could help to identify the
underlying mechanism of flavour symmetry breaking. 

In this paper we have presented a general 
analysis of the MFV effective theory in the 
$\Delta F=1$ sector. From the current stringent 
bounds on possible deviations from the SM in various 
$B$-physics observables we have derived 
the bounds on the scale of new-physics reported 
in Table~\ref{table:bounds2}. 
As can be seen, these bounds are
perfectly compatible with new dynamics in the 
TeV range. We recall that the in models where 
new particles contribute to FCNC processes 
only at the loop level,
the bounds on the particle masses are one
order of magnitude weaker with respect to the 
the bounds reported in Table~\ref{table:bounds2}:
$\Lambda \sim 4\pi m_{\rm NP}$.
Thus in weakly-interacting theories respcting the 
MFV hypothesis and with no tree-level FCNC, 
we could expect new particles well within the 
reach of the LHC.

Using the bounds on the effective operators, 
taking into account the correlations implied by the 
observables measured so far, we have derived 
a series of predictions for future 
high-statistics studies of flavour physics. 
This has allowed us to identify
observables which could still exhibit 
large deviations for the SM 
even under the pessimistic hypothesis of MFV.
The most interesting ones are the rare
decays $B_{s,d} \to \mu^+\mu^-$, rare $B$
and $K$ decays with a neutrino pair in the final 
state, and the FB asymmetry in $B \to K^* \ell^+\ell^-$
decays. The allowed parameter space for the 
latter is shown in Fig.~\ref{fig:afb}. 

Using the current bounds on MFV operators
we have also identified a series of stringent 
tests of this symmetry principle. The most interesting 
{\em negative predictions} are summarised in 
Table~\ref{tab:obs2}. If these predictions were falsified
by future experiments, we could unambiguously conclude that 
there exist new flavour symmetry-breaking structures 
beyond the SM Yukawa couplings. 
The effective theory allows us to obtain also some 
{\em positive predictions}, namely correlations 
among different observables  which could still
exhibit a deviation form the SM. The most interesting 
of such positive predictions is the correlation 
between  $B_{s} \to \mu^+\mu^-$ and  $B_{d} \to \mu^+\mu^-$
implied by Eq.~(\ref{eq:Bll_p}) and 
illustrated in Fig.~\ref{fig:bsmuta}. A clear evidence of 
physics beyond the SM in both channels, respecting
this correlation, would provide a strong support 
of the MFV hypothesis.

\section*{Acknowledgements}
We thank J. Charles, S. Descotes-Genon and U. Haisch for interesting discussions, and D. Guadagnoli for his comments on the manuscript. This work is  supported by the EU under contract
MTRN-CT-2006-035482 {\em Flavianet}.


\begin{thebibliography}{10}
\footnotesize{

\bibitem{D'Ambrosio:2002ex}
  G.~D'Ambrosio, G.~F.~Giudice, G.~Isidori and A.~Strumia,
  Nucl.\ Phys.\  B {\bf 645} (2002) 155
  [arXiv:hep-ph/0207036].

\bibitem{Chivukula:1987py}
  R.~S.~Chivukula and H.~Georgi,
  Phys.\ Lett.\  B {\bf 188} (1987) 99.

\bibitem{Buras:2000dm}
  A.~J.~Buras, {\em et al.}
  Phys.\ Lett.\  B {\bf 500} (2001) 161
  [arXiv:hep-ph/0007085].

\bibitem{others}
  A.~Ali and D.~London,
  Eur.\ Phys.\ J.\  C {\bf 9} (1999) 687
  [arXiv:hep-ph/9903535];
  A.~J.~Buras,
  Acta Phys.\ Polon.\  B {\bf 34} (2003) 5615
  [arXiv:hep-ph/0310208];
  B.~Grinstein, V.~Cirigliano, G.~Isidori and M.~B.~Wise,
  Nucl.\ Phys.\  B {\bf 763} (2007) 35
  [arXiv:hep-ph/0608123].



\bibitem{SUSY} 
  E.~Gabrielli and G.~F.~Giudice,
  Nucl.\ Phys.\  B {\bf 433}, 3 (1995)
  [Erratum-ibid.\  B {\bf 507}, 549 (1997)]
  [arXiv:hep-lat/9407029];
  L.~J.~Hall and L.~Randall,
  Phys.\ Rev.\ Lett.\  {\bf 65} (1990) 2939.


\bibitem{SUSYRGE}
  P.~Paradisi, M.~Ratz, R.~Schieren and C.~Simonetto,
  arXiv:0805.3989 [hep-ph];
  G.~Colangelo, E.~Nikolidakis and C.~Smith,
  arXiv:0807.0801 [hep-ph];
  B.~Dudley and C.~Kolda,
  arXiv:0805.4565 [hep-ph].

\bibitem{SUSYR}
 E.~Nikolidakis and C.~Smith,
  Phys.\ Rev.\  D {\bf 77}, 015021 (2008)
  [arXiv:0710.3129 [hep-ph]].


\bibitem{EDIM}
  G.~Cacciapaglia, C.~Csaki, J.~Galloway, G.~Marandella, J.~Terning and A.~Weiler,
  JHEP {\bf 0804} (2008) 006
  [arXiv:0709.1714 [hep-ph]].

\bibitem{UTfit}
  M.~Bona {\it et al.}  [UTfit Collaboration],
  JHEP {\bf 0803} (2008) 049
  [arXiv:0707.0636 [hep-ph]]; for UT angle $\gamma$ we use the updated value on \verb"http://www.utfit.org/".

\bibitem{HFAG}
  E.~Barberio {\it et al.}  [Heavy Flavor Averaging Group (HFAG)
                  Collaboration],
  arXiv:0704.3575 [hep-ex], and online update at 
  \verb"http://www.slac.stanford.edu/xorg/hfag"

\bibitem{Misiak:2006zs}
  M.~Misiak {\it et al.},
  Phys.\ Rev.\ Lett.\  {\bf 98}, 022002 (2007)
  [arXiv:hep-ph/0609232].

\bibitem{Misiak:2006ab}
  M.~Misiak and M.~Steinhauser,
  Nucl.\ Phys.\  B {\bf 764} (2007) 62
  [arXiv:hep-ph/0609241].

\bibitem{Haisch:2008ar}
  U.~Haisch,
  arXiv:0805.2141 [hep-ph].


\bibitem{Iwasaki:2005sy}
  M.~Iwasaki {\it et al.}  [Belle Collaboration],
  Phys.\ Rev.\  D {\bf 72}, 092005 (2005)
  [arXiv:hep-ex/0503044].

\bibitem{Aubert:2004it}
  B.~Aubert {\it et al.}  [BABAR Collaboration],
  Phys.\ Rev.\ Lett.\  {\bf 93}, 081802 (2004)
  [arXiv:hep-ex/0404006].

\bibitem{Bobeth:1999mk}
  C.~Bobeth, M.~Misiak and J.~Urban,
  Nucl.\ Phys.\  B {\bf 574}, 291 (2000)
  [arXiv:hep-ph/9910220].

\bibitem{Asatryan:2001zw}
  H.~H.~Asatryan, H.~M.~Asatrian, C.~Greub and M.~Walker,
  Phys.\ Rev.\  D {\bf 65} (2002) 074004
  [arXiv:hep-ph/0109140].

\bibitem{Asatrian:2002va}
  H.~M.~Asatrian, K.~Bieri, C.~Greub and A.~Hovhannisyan,
  Phys.\ Rev.\  D {\bf 66}, 094013 (2002)
  [arXiv:hep-ph/0209006].

\bibitem{Ghinculov:2003qd}
  A.~Ghinculov, T.~Hurth, G.~Isidori and Y.~P.~Yao,
  Nucl.\ Phys.\  B {\bf 685}, 351 (2004)
  [arXiv:hep-ph/0312128].

\bibitem{Bobeth:2003at}
  C.~Bobeth, P.~Gambino, M.~Gorbahn and U.~Haisch,
  JHEP {\bf 0404}, 071 (2004)
  [arXiv:hep-ph/0312090].

\bibitem{Huber:2005ig}
  T.~Huber, E.~Lunghi, M.~Misiak and D.~Wyler,
  Nucl.\ Phys.\  B {\bf 740}, 105 (2006)
  [arXiv:hep-ph/0512066].

\bibitem{Huber:2007vv}
  T.~Huber, T.~Hurth and E.~Lunghi,
  arXiv:0712.3009 [hep-ph].

\bibitem{Ligeti:2007sn}
  Z.~Ligeti and F.~J.~Tackmann,
  Phys.\ Lett.\  B {\bf 653} (2007) 404
  [arXiv:0707.1694 [hep-ph]].

\bibitem{:2007kv}
  T.~Aaltonen {\it et al.}  [CDF Collaboration],
  Phys.\ Rev.\ Lett.\  {\bf 100} (2008) 101802
  [arXiv:0712.1708 [hep-ex]].

\bibitem{Buchalla:1995vs}
  G.~Buchalla, A.~J.~Buras and M.~E.~Lautenbacher,
  Rev.\ Mod.\ Phys.\  {\bf 68} (1996) 1125
  [arXiv:hep-ph/9512380].

\bibitem{fbslattice}
  Q.~Mason {\it et al.}  [HPQCD Collaboration and UKQCD Collaboration],
  Phys.\ Rev.\ Lett.\  {\bf 95}, 052002 (2005)
  [arXiv:hep-lat/0503005];
  C.~Bernard {\it et al.}  [Fermilab Lattice, MILC and HPQCD Collaborations],
  PoS {\bf LATTICE2007}, 370 (2007).


\bibitem{babar:2008ju}
   {B. Aubert \it et al.}  [BABAR Collaboration],
  arXiv:0804.4412 [hep-ex]; Phys.\ Rev.\  D {\bf 73}, 092001 (2006)
  [arXiv:hep-ex/0604007].

\bibitem{Beneke:2001at}
  M.~Beneke, T.~Feldmann and D.~Seidel,
  Nucl.\ Phys.\  B {\bf 612}, 25 (2001)
  [arXiv:hep-ph/0106067].

\bibitem{Beneke:2004dp}
  M.~Beneke, T.~Feldmann and D.~Seidel,
  Eur.\ Phys.\ J.\  C {\bf 41} (2005) 173
  [arXiv:hep-ph/0412400].

\bibitem{Ali:2006ew}
  A.~Ali, G.~Kramer and G.~h.~Zhu,
  Eur.\ Phys.\ J.\  C {\bf 47} (2006) 625
  [arXiv:hep-ph/0601034].

\bibitem{Ball:2004rg}
  P.~Ball and R.~Zwicky,
  Phys.\ Rev.\  D {\bf 71} (2005) 014029
  [arXiv:hep-ph/0412079].

\bibitem{Adler:2008zz}
  S.~Adler {\it et al.}  [E787 Collaboration],
  Phys.\ Rev.\  D {\bf 77} (2008) 052003;
  V.~V.~Anisimovsky {\it et al.}  [E949 Collaboration],
  Phys.\ Rev.\ Lett.\  {\bf 93} (2004) 031801
  [arXiv:hep-ex/0403036].


\bibitem{Antonelli:2008jg}
  M.~Antonelli {\it et al.}  [FlaviaNet Working Group on Kaon Decays],
  arXiv:0801.1817 [hep-ph] and online update at 
  \verb"http://www.lnf.infn.it/wg/vus/" from Rare K decays and Decay Constants.


\bibitem{Brod:2008ss}
  J.~Brod and M.~Gorbahn,
  arXiv:0805.4119 [hep-ph].



\bibitem{Buras:2006gb}
  A.~J.~Buras, M.~Gorbahn, U.~Haisch and U.~Nierste,
  JHEP {\bf 0611}, 002 (2006)
  [arXiv:hep-ph/0603079];
  Phys.\ Rev.\ Lett.\  {\bf 95}, 261805 (2005)
  [arXiv:hep-ph/0508165].

  
\bibitem{Isidori:2005xm}
  G.~Isidori, F.~Mescia and C.~Smith,
  Nucl.\ Phys.\  B {\bf 718}, 319 (2005)
  [arXiv:hep-ph/0503107].

\bibitem{Mescia:2007kn}
  F.~Mescia and C.~Smith,
  Phys.\ Rev.\  D {\bf 76}, 034017 (2007)
  [arXiv:0705.2025 [hep-ph]].



\bibitem{Buras:2003td}
  A.~J.~Buras,
  Phys.\ Lett.\  B {\bf 566}, 115 (2003)
  [arXiv:hep-ph/0303060].

\bibitem{Bobeth:2005ck}
  C.~Bobeth, M.~Bona, A.~J.~Buras, T.~Ewerth, M.~Pierini, L.~Silvestrini and A.~Weiler,
  Nucl.\ Phys.\  B {\bf 726} (2005) 252
  [arXiv:hep-ph/0505110].

\bibitem{Haisch:2007ia}
  U.~Haisch and A.~Weiler,
  Phys.\ Rev.\  D {\bf 76}, 074027 (2007)
  [arXiv:0706.2054 [hep-ph]].

\bibitem{Babu:1999hn}
  K.~S.~Babu and C.~F.~Kolda,
  Phys.\ Rev.\ Lett.\  {\bf 84} (2000) 228
  [arXiv:hep-ph/9909476].

\bibitem{Isidori:2001fv}
  G.~Isidori and A.~Retico,
  JHEP {\bf 0111} (2001) 001
  [arXiv:hep-ph/0110121].

\bibitem{Altmannshofer:2007cs}
  W.~Altmannshofer, A.~J.~Buras and D.~Guadagnoli,
  JHEP {\bf 0711} (2007) 065
  [arXiv:hep-ph/0703200].



\bibitem{Gambino:2003zm}
  P.~Gambino, M.~Gorbahn and U.~Haisch,
  Nucl.\ Phys.\  B {\bf 673} (2003) 238
  [arXiv:hep-ph/0306079].

\bibitem{Gorbahn:2004my}
  M.~Gorbahn and U.~Haisch,
  Nucl.\ Phys.\  B {\bf 713}, 291 (2005)
  [arXiv:hep-ph/0411071].





\bibitem{BIR}
  G.~Buchalla, G.~Isidori and S.~J.~Rey,
  Nucl.\ Phys.\  B {\bf 511} (1998) 594
  [arXiv:hep-ph/9705253].

\bibitem{Neubertlee}
  S.~J.~Lee, M.~Neubert and G.~Paz,
  Phys.\ Rev.\  D {\bf 75} (2007) 114005
  [arXiv:hep-ph/0609224].


\bibitem{Hubernew}
  T.~Huber, T.~Hurth and E.~Lunghi,
  arXiv:0807.1940 [hep-ph].






\bibitem{Ball:2001fp}
  P.~Ball and R.~Zwicky,
  JHEP {\bf 0110} (2001) 019
  [arXiv:hep-ph/0110115].

\bibitem{Ali:1999mm}
  A.~Ali, P.~Ball, L.~T.~Handoko and G.~Hiller,
  Phys.\ Rev.\  D {\bf 61} (2000) 074024
  [arXiv:hep-ph/9910221].

\bibitem{Hiller:2003js}
  G.~Hiller and F.~Kruger,
  Phys.\ Rev.\  D {\bf 69}, 074020 (2004)
  [arXiv:hep-ph/0310219].

\bibitem{Hill:2004if}
  R.~J.~Hill, T.~Becher, S.~J.~Lee and M.~Neubert,
  JHEP {\bf 0407} (2004) 081
  [arXiv:hep-ph/0404217].

\bibitem{Egede:2008uy}
  U.~Egede, T.~Hurth, J.~Matias, M.~Ramon and W.~Reece,
  arXiv:0807.2589 [hep-ph].

\bibitem{Ball:2004ye}
  P.~Ball and R.~Zwicky,
  Phys.\ Rev.\  D {\bf 71} (2005) 014015
  [arXiv:hep-ph/0406232].

\bibitem{Buchalla:2000sk}
  G.~Buchalla, G.~Hiller and G.~Isidori,
  Phys.\ Rev.\  D {\bf 63} (2001) 014015
  [arXiv:hep-ph/0006136].

\bibitem{Colangelo:2006vm}
  P.~Colangelo, F.~De Fazio, R.~Ferrandes and T.~N.~Pham,
  Phys.\ Rev.\  D {\bf 73} (2006) 115006
  [arXiv:hep-ph/0604029].

\bibitem{Misiak:private}
  M.~Misiak, private communication.




\bibitem{PDG08}
  W.~M.~Yao {\it et al.}  [Particle Data Group],
  J.\ Phys.\ G {\bf 33}, 1 (2006), and online update at  
  \verb"http://pdg.lbl.gov/"

\bibitem{Tantalo}
  N.~Tantalo,
  arXiv:hep-ph/0703241;S. Hashimoto, 
  hep-ph/0411126; D. Becirevic, hep-ph/0310072.

\bibitem{Gambino:2004mv}
  P.~Gambino, U.~Haisch and M.~Misiak,
  Phys.\ Rev.\ Lett.\  {\bf 94}, 061803 (2005)
  [arXiv:hep-ph/0410155].

\bibitem{Ishikawa:2006fh}
  A.~Ishikawa {\it et al.},
  Phys.\ Rev.\ Lett.\  {\bf 96} (2006) 251801
  [arXiv:hep-ex/0603018].

\bibitem{Abe:2004ir}
  K.~Abe {\it et al.}  [Belle Collaboration],
  arXiv:hep-ex/0410006.

\bibitem{Aubert:2008ps}
  B.~Aubert  [The BABAR Collaboration],
  arXiv:0807.4119 [hep-ex].

\bibitem{Bobeth:2007dw}
  C.~Bobeth, G.~Hiller and G.~Piranishvili,
  JHEP {\bf 0712} (2007) 040
  [arXiv:0709.4174 [hep-ph]].

\bibitem{:2007zk}
  K.~F.~Chen {\it et al.}  [BELLE Collaboration],
  Phys.\ Rev.\ Lett.\  {\bf 99} (2007) 221802
  [arXiv:0707.0138 [hep-ex]];
  B.~Aubert {\it et al.}  [BABAR Collaboration],
  Phys.\ Rev.\ Lett.\  {\bf 94} (2005) 101801
  [arXiv:hep-ex/0411061].

\bibitem{Nix:2007sa}
  J.~Nix {\it et al.}  [E391a Collaboration],
  Phys.\ Rev.\  D {\bf 76} (2007) 011101
  [arXiv:hep-ex/0701058].
}
\end{thebibliography}
\end{document}